\documentclass[fleqn,usenatbib]{mnras}
\usepackage{graphicx}
\usepackage{txfonts}
\usepackage{gensymb}
\newcommand{\Msun}{M$_{\sun}$}
\bibpunct[,]{(}{)}{;}{a}{,}{,}

\usepackage{xcolor}
 
\begin{document}

\title[Chemical evolution within and outside of spiral arms]
{2D-Galactic chemical evolution: the role of the spiral density wave}

\author[Moll\'{a} et al. ]
{M. Moll\'{a}$^{1}$\thanks{E-mail:mercedes.molla@ciemat.es}, S. Wekesa$^{2}$, O. Cavichia$^{3}$, {\'A}. I. D\'{\i}az$^{4,5}$, 
B. K. Gibson$^{6,10}$, 
\newauthor 
F. F. Rosales-Ortega$^{7}$, Y. Ascasibar$^{4,5}$, 
D. S. Wamalwa$^{8}$, S. F. S\'{a}nchez$^{9}$ \\
$^{1}$ Departamento de Investigaci\'{o}n B\'{a}sica, CIEMAT, Avda. Complutense 40, E-28040 Madrid, Spain\\ 
$^{2}$ University of Nairobi, Nairobi, Kenya\\ 
$^{3}$ Instituto de F{\'i}sica e Qu{\'i}mica, Universidade Federal de Itajub{\'a}, Av. BPS, 1303, 37500-903, Itajub{\'a}-MG, Brazil\\
$^{4}$Universidad Aut\'{o}noma de Madrid, E-28049 Madrid, Spain \\
$^{5}$ Astro-UAM, Unidad Asociada CSIC, Universidad Aut\'{o}noma de Madrid, 28049, Madrid, Spain\\
$^{6}$ E.~A. Milne Centre for Astrophysics, University of Hull, Hull, HU6~7RX, United Kingdom\\
$^{7}$ Instituto Nacional de Astrof\'{\i}sica, \'{O}ptica y Electr\'{o}nica, Luis E. Erro 1, 72840 Tonantzintla, PUE, M\'{e}xico\\
$^{8}$ Meru University of Science \& Technology, Meru, Kenya\\
$^{9}$ Instituto de Astronom\'{\i}a, Universidad Nacional Aut\'{o}noma de M\'{e}xico, A.P. 70-264, 04510, M\'{e}xico, D.F.\\
$^{10}$ Joint Institute for Nuclear Astrophysics - Center for the Evolution of the Elements (JINA-CEE)}
\pagerange{\pageref{firstpage}--\pageref{lastpage}} 
\pubyear{2019}
\maketitle 
\label{firstpage}

\begin{abstract}
We present a 2-dimensional chemical evolution code applied to a Milky Way type galaxy, incorporating the role of spiral arms in shaping azimuthal abundance variations, and confront the predicted behaviour with recent observations taken with integral field units. To the usual radial distribution of mass, we add the surface density of the spiral wave and study its effect on star formation and elemental abundances. We compute five different models: one with azimuthal symmetry which depends only on radius, while the other four are subjected to the effect of a spiral density wave. At early times, the imprint of the spiral density wave is carried by both the stellar and star formation surface densities; conversely, the elemental abundance pattern is less affected. At later epochs, however, differences among the models are diluted, becoming almost indistinguishable given current observational uncertainties. At the present time, the largest differences appear in the star formation rate and/or in the outer disc (R$\ge$ 18\,kpc). The predicted azimuthal oxygen abundance patterns for $t \le 2$\,Gyr are in reasonable agreement with recent observations obtained with VLT/MUSE for NGC~6754.
 \end{abstract}


\section{Introduction}

The nebular emission arising from extragalactic H{\sc ii} regions has
played an important role in our understanding of the chemical
evolution in spiral and irregular galaxies. Nebular emission lines
from individual H{\sc ii} regions have been historically the main tool
at our disposal for the direct measurement of the gas phase abundance
at discrete spatial positions in low-redshift galaxies. They trace the
young massive star component in galaxies, illuminating and ionising
volumes of the interstellar medium (ISM), but also bear chemical imprint
of the integrated star formation of the galaxy. H{\sc ii} regions emit 
forbidden lines from a variety of
heavy elements, with which, using appropriate constraints on physical
conditions, ionic and elemental abundances can be derived. 
Most of the observations targeting
nebular emission have been made with single--aperture or long--slit
spectrographs, resulting in samples of typically a dozen or fewer
H{\sc ii} regions per galaxy
\citep{pag79,mcall85,skil89,diaz89,zar94,mr94,vzee98,gar02,gav04,mous06,arls10,ros11,bre12}
or single spectra of large survey samples \citep{trem04} for SDSS.
Regardless of approach, the overwhelming result from these studies is
that spiral galaxies present clear negative radial abundance gradients.

Clearly, galaxies are complex systems, not fully represented by a
singular spectrum or a simple azimuthally-averaged radial distribution of 
spectral emission
line intensities in one dimension (1D).  Integral Field Spectroscopy
(IFS) is a technique that affords the opportunity to obtain spectra of 
extended sources and/or multiple regions
as a function of spatial position within a given system. For example, 
IFS instruments with large
fields--of--view, such as PPAK \citep{sanchez12a}, SAMI \citep{croom12},
SDSS-IFU \citep{bundy15}, and MUSE \citep{san15b}, allow one to undertake 
full two--dimensional (2D) sampling of nearby galaxies, 
instead of relying upon
single--aperture spectra; this then provides access to data relevant
fundamental issues in galactic structure and evolution, including 
spatially-resolved star formation history (SFH)--interstellar medium (ISM) coupling 
\citep{ros11,sanchez12b}.  In particular, CALIFA has
demonstrated a local downsizing to form disc galaxies, with an
inside-out assembly process in the growth of their discs \citep{per13,cf14},
and the existence of a characteristic common radial gradient,
independent of galaxy properties
\citep{sanchez12b,san14,sanmen16b,sanmen18}.

Chemical evolution models (CEM) are the usual tool to interpret 
elemental abundance patterns. However, CEMs commonly assume that abundance
distributions are azimuthally symmetric, and ignore the potential
dispersion in abundances at a given position. In order to gain a deeper
insight into the mechanisms that shape the chemical evolution of
galaxies and take advantage of the newest (and next) generation of 
IFS survey data, we require, on the
theoretical side, powerful and robust modelling of the physical
processes and the evolution of the gas phase and stellar populations
of a galaxy; from these, we could link for the first time real 2D
observations with appropriate and complementary theoretical models, in order to
progress from the customary, azimuthally--averaged, radial
representation of galaxy observable to a spatially--resolved
understanding.  This synergy between high--quality IFS observations and
2D theoretical modelling represents a step forward in our understanding 
of the physics of galaxy evolution.

Modelling the distribution of elements within a spiral disc 
requires, at the least, a 2D description of the arms, bars and other
structures produced by mergers or interactions. Such a description
is the thrust of our current work; in particular, our initial 
focus is the potential differences that might exist between arm and
inter-arm regions due to the influence of these aforementioned
2D structures. Spiral arms are typical features of late-type disc galaxies,
which make up roughly 70$\% $ of the bright galaxies in the local
volume. They are found not only in the distribution of cold gas and
young bright stars, but also in the old stellar populations
\citep{rix95}. One of our expectations is that a spiral wave could
modify the star formation rate (hereafter, SFR), since the presence
of arms increases the surface density of gas within them and, 
consequently, could result in enhanced star formation (with associated 
enhanced local feedback). Moreover, spiral arms
may be more direct triggers of star formation since they may raise the
probability of cloud-cloud collisions and/or increase the 
molecular cloud frequency, through shocks driven as the ambient ISM
enters the arm \citep{kob07,dib09}. However, although star formation 
is stimulated by
galactic shocks due to the spiral wave, near the co-rotation
resonance, there has been an observed gap in the gas density
\citep{amores09}, which may be associated with a drastic reduction of
the SFR. These results are also in agreement with the studies of
\citet{dobbs11} who suggested that spiral arms are mainly 'organized'
features whose main effect on ISM is to delay and crowd the gas.

From an observational perspective, only a few works have addressed
the arm-interarm (or azimuthal variation) abundance question
\citep{mar96,cedres02,ryder05,ros11,cedres12,li13}, most of them without
finding any difference between zones, except for \citet{cedres12},
who found rich metallicity knots located in the arms, as compared to
other H{\sc ii} regions in the discs of NGC~628 and NGC~6946. 
\citet{sanmen16a} have studied the abundances of NGC~6754, finding
residual abundances between arm and interarm regions, the distribution
of these residuals being positive at the trailing side of the arm and
negative at the leading edge, with a total amplitude of $\sim
0.1$\,dex. More recently, \citet{sanmen17} analysed a sample of CALIFA
galaxies, computing the radial gradient of oxygen for arm and interarm
regions, finding subtle differences, yet statistically significant, at
least in flocculent galaxies. These findings are supported by other
works \citep{ho17,ho18} which estimate systematic azimuthal
variations in the oxygen abundances of NGC~1365 and NGC~2997,
respectively, of the order of $\sim$0.06 to 0.2~dex, over a wide
radial range of 0.3--0.7~$R_{25}$, peaking at the two spiral
arms. Also, \citet{vogt17} from MUSE observations in the galaxy HCG~
91c, measure azimuthal variations, with spiral arms more metal-rich
than interarm regions.

From a theoretical perspective, most works treating this point
analyse hydro-dynamical simulations from the dynamical point of
view. \citet{baba15} investigated, using a three dimensional N-body
hydro-dynamical simulation, the evolution of grand design spiral
arms, finding that the arms are not stationary, but 
rather self-excited dynamic patterns, changing within a few
hundred millions years. Only some works have studied the metallicity
or abundance variations by taking into account the existence of the
spiral arm. \citet{grand16} analyse a cosmological simulation realised at
high resolution for a Milky Way-sized halo, demonstrating the existence of
azimuthal variations between the arm and the interarm regions in the
stellar metallicities (more metal rich in the trailing edge and more
metal poor in the leading one). Only limited work to date has treated CEM in
2D, such as  \citet{acharova13}. Very recently, almost simultaneous to
this work, \citet{spi19} have developed a 2D CEM, taking into account
the azimuthal surface density variations and the 1D CEM code of
\citet{mf89}, to study azimuthal differences in the oxygen
abundances. These authors separate the study of the density
fluctuations obtained from the chemo-dynamical model of \citet{min13}
from these ones due to spiral arms, to discriminate both effects 
upon elemental abundances. They find
variations of the order of $\sim$ 0.1\,dex, being more evident in the outer
regions of the disc, when the density fluctuations are
included. With the spiral arm alone they produce smaller 
differences, except in the corotation region. 
By changing the spiral wave parameters, e.g., by modifying the spiral
wave pattern, which moves the corotation region towards outer radii, 
or assuming only one spiral arm, differences increase. These variations were
higher in the early times, erasing with time.

In summary, our objective is to develop comprehensive and
sophisticated 2D chemical evolution models for spiral and irregular
galaxies, in order to constrain the observed IFS data. To
accomplish this, we first convert the existing
one-dimensional (1D) chemical evolution model, detailed in
\citep{fer92,fer94,md05,molla14}, into a two dimensional (2D) model,
with a spatial resolution that can be adjusted to the new two
dimensional abundance distributions provided by current
databases. This new code includes the typical structures observed
in a spiral galaxy, including the bulge, the stellar bar, and the spiral
wave. Our basic framework is based on the {\sc MULCHEM} updated
models presented in \citet{mol15,mol16,mol17,mol19}, each of which have been
calibrated to fit the Milky Way Galaxy (MWG), and
are also valid for galaxies of a wide range of dynamical mass
(Moll{\'a} et al. in preparation).

We have divided our project into three phases: 
\begin{enumerate}
\item As a first step, a stationary spiral wave is added at the
  initial time $t=0$ as an over-density with respect to the
  exponentially decreasing radial surface density distribution of the
  disc.  In principle, arms do change with time, but it is not clear
  whether they are long or short-lived features. At present, there are
  some models that show that the density structure becomes stationary
  after integration times between 400 and 1200\,Myr \citep{ant11}.
  Moreover, even if the assumption of a stationary arm is not entirely
  realistic, one may expect that it should yield the strongest
  arm-interarm contrast.  It is our main objective to test whether
  such perturbations may induce detectable differences (that is, larger
  than the typical uncertainties of the data) in the SFR and elemental
  abundances, and thus we will consider this first check as a
  prototype model to evaluate its potential impact upon the observable
  properties of the disc.
  
  We then include the rotation of the arm in our model and
  compare with the results obtained for a stationary wave. Any mixing,
  caused by the wave rotation and/or any movement of gas between
  spaxels or spatial cells, is expected to produce a larger dispersion
  of chemical abundances, and therefore a likely decrease in the arm-interarm
  contrast, compared to the stationary model.  We also show a model
  where the overdensity is added to the galactic halo, so that the
  growth of the spiral arm takes place smoothly with time,
  simultaneous to the disc formation, and therefore much smaller
  differences with respect to the unperturbed model are expected.

\item In our subsequent work, a parametric study will be carried out,
  where we will compute a more complete set of models, varying the
  parameters that define the spiral wave, in order to check which
  quantities have a more significant impact upon the physical
  properties of the galaxies.  Within this new set of models, the 
  possibility of creating
  the spiral arm at different moments in a disc's evolution will 
  be explored.

\item Finally, we will take into account the time-dependent nature of 
  these structures 
  \citep[see Fig.~2 from][]{roc13}. The
  time evolution of the spiral arm, including the effect of non–axisymmetric
  co–planar radial flows driven by the stellar bar, as shown by
  \citet{cavichia14}, implies that we have to take into account the
  movement of mass from one spaxel to its neighbours. The variety of 
  features and the nature of our framework results in a complex 
  extension to the base model presented here, and lie beyond the 
  scope of this initial work; this does not undermine the results of
  the current paper. Nevertheless, it will form the subject of our 
  currently 'in prep' work.

\end{enumerate}

In nutshell, we outline here how to modify our basic 1D framework, and 
show the first models applied to a Milky Way-type
galaxy (MWG). We then analyse the effects of the over-density created by a
spiral wave on the resulting elemental abundances and SFR in
predicted 2D distributions. In Section~\ref{modelos}, we describe the 
adopted methodology and, specifically, how the arms are included. In
Section~\ref{results}, we show the results for a set of computed
models, comparing them with the ones obtained with a classical
azimuthally symmetric 1D model, mainly in relation to the resulting differences in the
distributions of elemental abundances and the SFR. We also confront
these results with both extant data and comparable models. Our conclusions
are given in Section~\ref{con}.

\section{Models}
\label{modelos}
\subsection{The azimuthally symmetric scenario}

Starting from the parametric 1D framework of \citet{md05} and
\citet{molla14}, we apply the multiphase chemical
 {\sc MulChem} model described by \citet[][ hereinafter MOL15,
  MOL16, MOL17, and MOL19, respectively]{mol15, mol16,mol17,mol19}, to
the Milky Way Galaxy (MWG), in order to develop a new generation of
bi-dimensional {\sc 2D-MulChem} models.

The azimuthally symmetric model for the MWG starts with a spherical
protohalo with a dynamical mass $M_{\rm dyn}=10^{12}{\rm M}_{\sun}$ \citep{piffl14,watkins19}. 
The initial mass of the halo collapses onto the equatorial plane to form
a disc with mass $M_{\rm D}$. We use the equations from \citet{sal07}
giving the rotation curves for both halo and disc as a function of the
dynamical mass of the dark matter halo, and, with them, we calculate
the corresponding radial mass distributions for a galactic halo and
disc within a radius $R$: $M_{\textrm{\scriptsize H}}(R)$ and
$M_{\textrm{\scriptsize disc}}(R)$. Moreover, we have included 
in the term {\sl D} the mass of the bulge:
$M_{\textrm{\scriptsize D}}(R)=M_{\textrm{\scriptsize disc}}(R)+M_{\textrm{\scriptsize B}}(R)$, 
which we assume to be zero, that is, $M_{\textrm{\scriptsize D}}(R)=M_{\textrm{\scriptsize  disc}}(R)$, for radial regions located at $R>R_{\textrm{\scriptsize B}}=4$\,kpc \citep[see][ for details concerning
  the calculation of these mass distributions]{mol16}.
\begin{figure*}
\includegraphics[width=0.48\textwidth,angle=0]{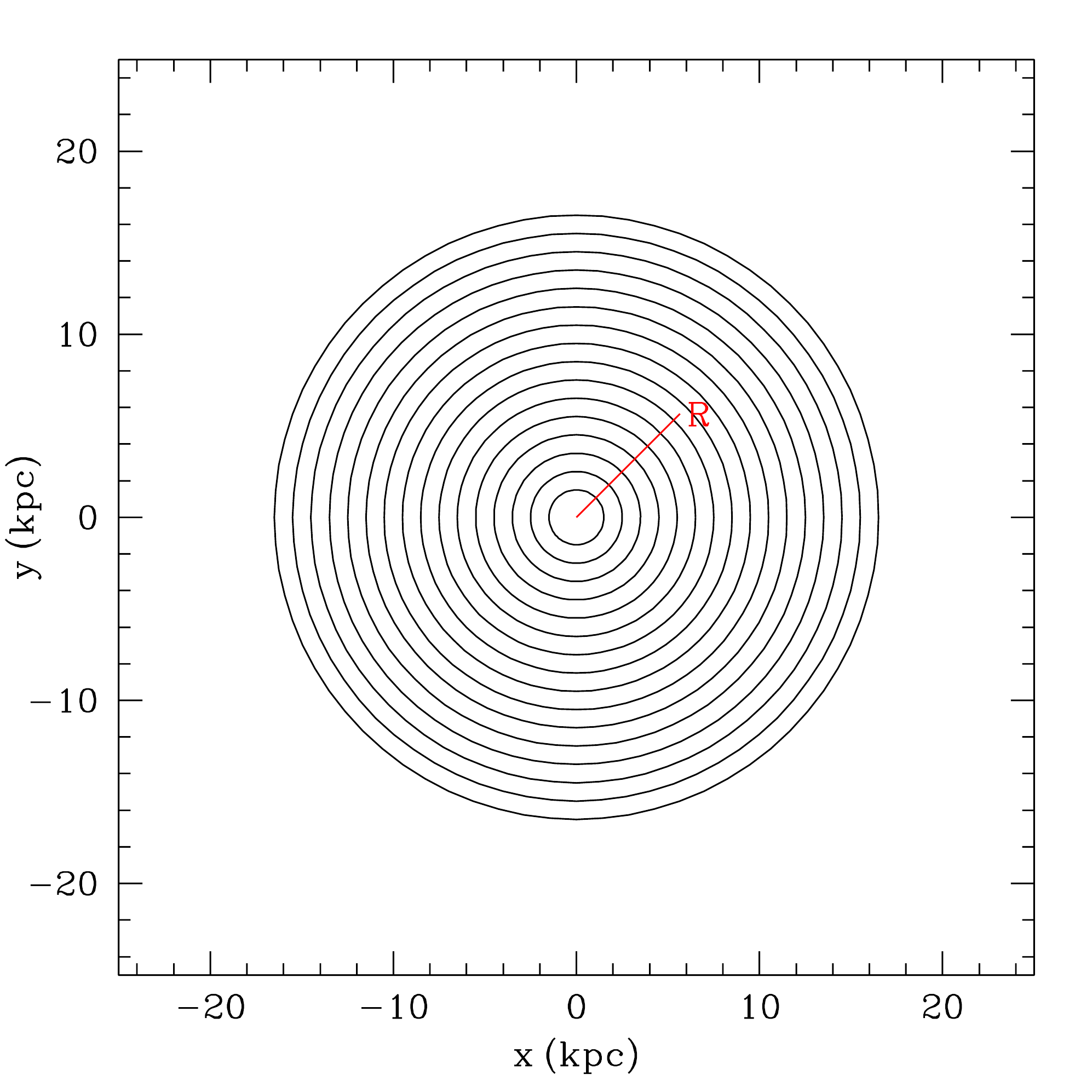}
\includegraphics[width=0.48\textwidth,angle=0]{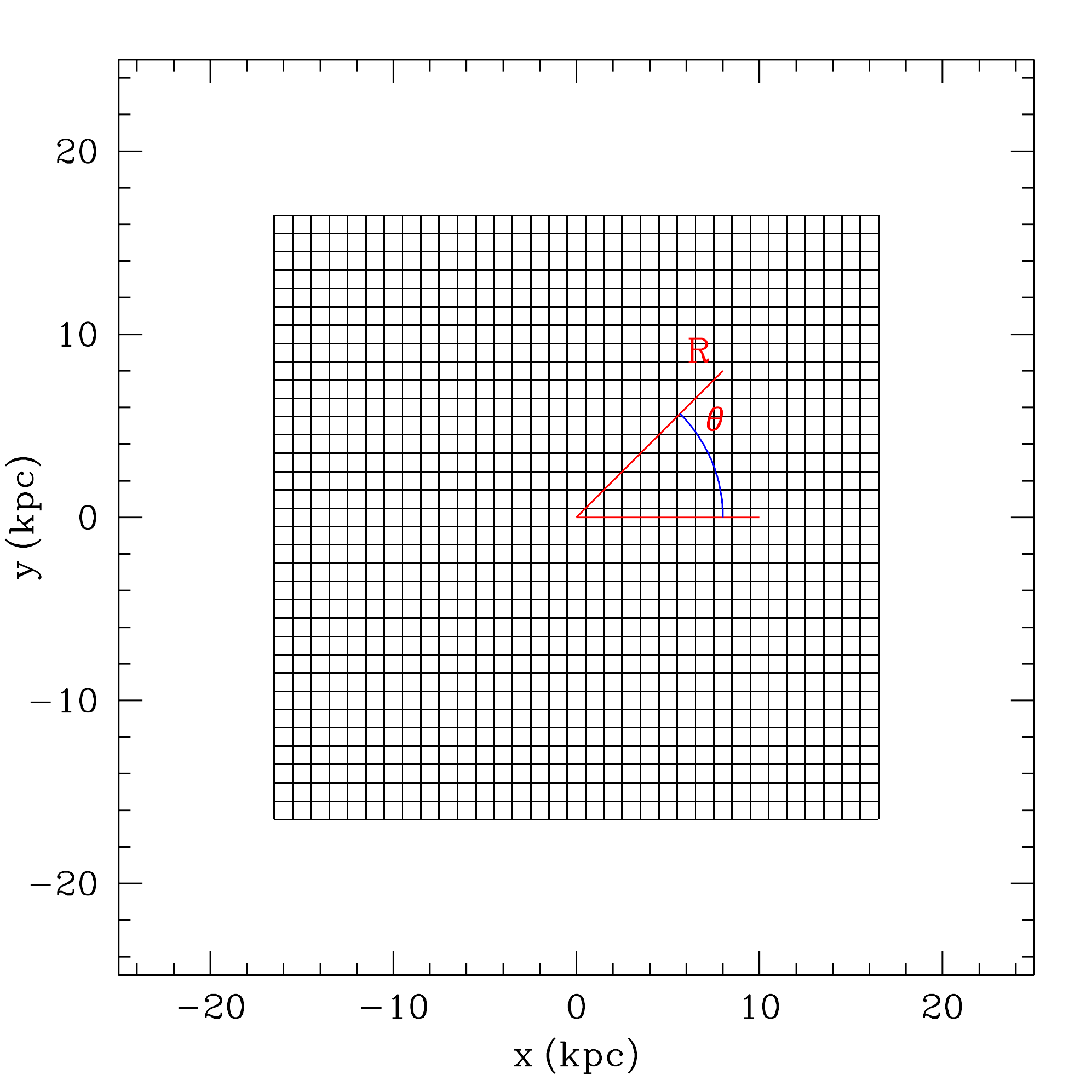}
\caption{Coordinate representations: a) Radial {\sc MulChem} (1D) scheme, depending only on the galactocentric distance $R$. b) Cartesian grid used in 2D-{\sc Mulchem}, where each region is centred on a point $(x, y)$.}
\label{scheme}
\end{figure*}
\begin{table*}[h]
\caption{Radial distributions of total mass, present-day disc, bulge, and halo mass, collapse time scale, and total and disc rotation velocities for radial regions in a MWG-like galaxy.}
\begin{center}
\begin{tabular}{rrrrrrrr}
\hline
$R$ & $\Delta M(R)$ & $\Delta M_{\textrm{\scriptsize disc}}(R)$ & $\Delta M_{\textrm{\scriptsize B}}(R)$  &  $\Delta M_{\textrm{\scriptsize H}}(R)$ & $\tau(R)$  & $V_{\textrm{\scriptsize rot,tot}}$  & $V_{\textrm{\scriptsize rot,D}}$  \\
$[\textrm kpc]$  &   [$10^{9}$\,\Msun]      &   [$10^{9}$\,\Msun]         &   [$10^{9}$\,\Msun]          &   [$10^{9}$\,\Msun]     &  [Gyr]  & [$km\,s^{-1}$] & [$km\,s^{-1}$]\\
\hline
  0.00   &    6.57   &    0.00   &    6.46   &    0.000  &     1.95 &    0.00 &    0.00 \\ 
 1.00   &    3.65   &    1.10   &    2.52   &    0.028  &     2.72 &   69.99 &   69.12 \\ 
 2.00   &    6.11   &    4.40   &    1.53   &    0.189  &     3.80 &  111.90 &  109.79 \\ 
 3.00   &    8.55   &    7.29   &    0.778  &    0.484  &     4.60 &  140.87 &  137.25 \\ 
 4.00   &    10.3   &    9.08   &    0.364  &    0.884  &     5.37 &  161.34 &  155.96 \\ 
 5.00   &    11.2   &    9.81   &    0.000  &    1.360  &     6.27 &  175.80 &  168.42 \\ 
 6.00   &    11.6   &    9.72   &    0.000  &    1.900  &     7.28 &  185.88 &  176.27 \\ 
 7.00   &    11.5   &    9.07   &    0.000  &    2.460  &     8.54 &  192.74 &  180.69 \\ 
 8.00   &    11.1   &    8.09   &    0.000  &    3.030  &     10.2 &  197.21 &  182.54 \\ 
 9.00   &    10.5   &    6.94   &    0.000  &    3.600  &     12.3 &  199.93 &  182.47 \\ 
10.00   &    9.92   &    5.77   &    0.000  &    4.160  &     15.2 &  201.38 &  180.98 \\ 
11.00   &    9.31   &    4.64   &    0.000  &    4.680  &     19.2 &  201.93 &  178.46 \\ 
12.00   &    8.77   &    3.60   &    0.000  &    5.170  &     25.0 &  201.84 &  175.22 \\ 
13.00   &    8.31   &    2.69   &    0.000  &    5.620  &     33.7 &  201.32 &  171.47 \\ 
14.00   &    7.94   &    1.91   &    0.000  &    6.030  &     47.9 &  200.54 &  167.41 \\ 
15.00   &    7.66   &    1.26   &    0.000  &    6.400  &     73.3 &  199.61 &  163.18 \\ 
16.00   &    7.46   &    0.74   &    0.000  &    6.730  &     127 &  198.60 &  158.88 \\ 
17.00   &    7.33   &    0.41   &    0.000  &    7.010  &     221 &  197.59 &  154.58 \\ 
18.00   &    7.25   &    0.25   &    0.000  &    7.260  &     385 &  196.60 &  150.36 \\ 
19.00   &    7.22   &    0.15   &    0.000  &    7.470  &     669 &  195.66 &  146.25 \\ 
20.00   &    7.22   &    0.09   &    0.000  &    7.660  &     1160 &  194.79 &  142.28 \\ 
21.00   &    7.25   &    0.06   &    0.000  &    7.810  &     2020 &  193.99 &  138.48 \\ 
22.00   &    7.29   &    0.04   &    0.000  &    7.930  &     3510 &  193.27 &  134.84 \\ 
23.00   &    7.35   &    0.02   &    0.000  &    8.030  &     6100 &  192.63 &  131.39 \\ 
24.00   &    7.41   &    0.01   &    0.000  &    8.110  &     10600 &  192.06 &  128.11 \\ 
 \hline
\end{tabular}
\end{center}
\label{mwg-mass-R}
\end{table*}

In order to progress from 1D to 2D modelling, we must define our reference frame in terms of cylindrical ($R, \theta$) or, equivalently, cartesian ($x$,$y$) coordinates rather than radial coordinate $R$ only.
Panel (a) from Figure~\ref{scheme} shows our previous 1D scheme, where the galaxy was divided into concentric annuli (cylinders) between galactocentric radii $R \pm \frac{\Delta R}{2}$, with $\Delta R = 1$~kpc.
Each of these regions contains a mass $\Delta M(R) = M(R+\frac{\Delta R}{2}) - M(R-\frac{\Delta R}{2})$ distributed over an area $2\pi\,R\,\Delta R$.
Now, we define a plane with 33 regions between $x=-16$\,kpc and $x=+16$\,kpc and $y=-16$\,kpc and $y=+16$\,kpc.
Each region is therefore, $\Delta x \times \Delta y = 1\,\textrm{kpc}\times 1\,\textrm{kpc}$ wide, centred on $(x,y)$, or, equivalently:
\begin{eqnarray}
R&=& \sqrt{x^{2}+y^{2}}\\
\theta&=&\arctan{y/x}
\nonumber
\end{eqnarray}
as illustrated by Panel (b) in Fig.~\ref{scheme}.

At the initial time $t=0$, all the mass in the halo is assumed to be in form of gas, while $M_{\textrm{\scriptsize disc}}(R)=M_{\textrm{\scriptsize B}}(R)=0$. 
The infall of the gas from the halo to the disc or bulge is assumed to occur with a collapse time scale $\tau(R)$:
\begin{equation}
\tau(R)=-\frac{13.2}{\ln{\left(1-\frac{\Delta M_{\textrm{\scriptsize D}}(R,t_{\rm pr})}{\Delta M_{\textrm{\scriptsize H}}(R,0)}\right)}}\,[\mbox{Gyr}],
\label{eq_infall_rate}
\end{equation}
where $\Delta M_{\textrm{\scriptsize H}}(R,0)$ and $\Delta
M_{\textrm{\scriptsize D}}(R,t_{\rm pr})$ are, respectively, the initial
halo and disc masses at the end of the evolution (i.e., at the
present time $t=t_{\rm pr}$, which we assume to be 13.2~Gyr) in each
radial region located at galactocentric distance $R$. The infall rates
  resulting from these equations for galaxies with masses different
  from that of the MWG are provided in \citet{mol16}. There, they are
  compared with both empirical data and cosmological simulations, analysing the
  radial differences and the evolution with redshift.

These radial distributions are listed in Table~\ref{mwg-mass-R} where
we give, for each galactocentric distance $R$ in column (1), the total
mass of the protohalo $\Delta M(R)$, or initial mass of the halo,
$\Delta M_{\textrm{\scriptsize H}}(R,t=0)$, in column (2), the
expected final mass after the evolution, for disc, $\Delta
M_{\textrm{\scriptsize disc}}(R,t_{\rm pr})$, bulge, $\Delta
M_{\textrm{\scriptsize B}}(R,t_{\rm pr})$, and halo, $\Delta
M_{\textrm{\scriptsize H}}(R,t_{\rm pr})$, in columns 3 to 5, and the
necessary collapse time scale $\tau(R)$, in Gyr, to obtain these
distributions at the present time in column 6. In columns 7 and 8, the total rotation velocity and the corresponding one for the disc component are given.

To set up the initial masses in {\sc 2D-MulChem}, we use the same radial mass distribution $M_{\textrm{\scriptsize H}}(R,t=0)$ assumed in our {\sl classical scenario}. We then calculate the original halo density in a region located at a radius R:
\begin{equation}
\rho_{\textrm{\scriptsize{H}}}(R,t=0)=\frac{\Delta M_{\textrm{\scriptsize H}}(R,t=0)}{(2\pi R)h(R)}\,[\rm M_{\sun}\,pc^{-3}],
\end{equation}
where $h(R)$ is the height of each halo region between two cylinders and $\Delta M_{\textrm{\scriptsize H}}(R, t=0)=\Delta M(R)$ is the mass given in column 2 of Table~\ref{mwg-mass-R}. We assume now that the new halo regions are square prism with base $\Delta x\times \Delta y= 1\,kpc\times1\,kpc$, centred on the point $x,y$ or $R,\theta$, and that the density is the same as before. Thus, the mass in the halo corresponding to each cell is:
\begin{eqnarray}
\Delta M_{\textrm{\scriptsize H}}(R,\theta,t=0)& = & \Delta M_{\textrm{\scriptsize H}}(x,y,t=0)= \nonumber\\
\rho_{\textrm{\scriptsize H}}(R,t=0)\times h(R)\times \Delta x\times \Delta y & = & \Delta M(R)\frac{10^{6}}{(2\pi R)}\,{\rm M}_{\sun}
\label{mhalo}
\end{eqnarray}

Once the mass in each cell $(x,y)$ is computed, we assume that this gas falls onto the equatorial plane at the same collapse time $\tau(R)$ specified in Table~\ref{mwg-mass-R}. That is, all cells with the same $R$ will have the same initial mass and collapse time, without any dependence on the angle $\theta$.

\subsection{Molecular clouds and star formation}
\label{sfr}
As explained in \citet{md05}, we assume that star formation (SF) follows a Schmidt law in the halo regions; however, in the disc, the star formation occurs in two steps: first, molecular clouds, $c{\textrm{\scriptsize D}}$, form from diffuse gas, $g_{\textrm{\scriptsize D}}$; then, stars (of two mass ranges, low mass stars, $s_{1}$, and intermediate mass and massive stars, $s_{2}$), form through cloud-cloud collisions; a second star formation process then appears, resulting from the interaction of massive stars $s_{\textrm{\scriptsize {D,2}}}$  with the molecular clouds $c$ surrounding them. Therefore,  we have different processes defined in the galaxy:
\begin{enumerate}
\item Star formation by spontaneous fragmentation of gas in the halo:
$\propto \kappa_{h;1,2}\,g_{\textrm{\scriptsize H}}^{n}$, where we use $n = 1.5$, $g_{\textrm{\scriptsize H}}$ being the mass of (total) gas in each halo region.
\item disc formation by gas accretion from the halo or protogalaxy: $f\,g_{\textrm{\scriptsize H}}$, $f$ being a parameter such that $f(R)=1/\tau(R)$ (see Eq.~2)
\item Formation of molecular clouds, $c$, in the disc from the diffuse (atomic) gas, $g_{\textrm{\scriptsize D}}$. This process is proportional to the mass of diffuse gas:  $\propto \kappa_{c} g_{\textrm{\scriptsize D}}$.  
\item Star formation in the disc due to cloud-cloud collisions: $\propto \kappa_{s;1,2}c^{2}$, with $c$ in solar masses.
\item Diffuse gas restitution in the disc due to cloud-cloud collisions: $\propto \kappa_{s}'c^{2}$.
\item Induced star formation in the disc due to the interaction between molecular clouds, $c$, and massive stars, $s_{\textrm{\scriptsize{2,D}}}$: $\propto \kappa_{a;1,2}c\,s_{\textrm{\scriptsize {2,D}}}$
\item Diffuse gas restitution in the disc due to the induced star formation. Massive stars, $s_{\textrm{\scriptsize{2,D}}}$, induce SF in the surrounding molecular clouds, $c$, but their radiation also destroys a proportion of them: $\kappa_{a}'c\,s_{\textrm{\scriptsize{2,D}}}$\item Ejection of gas to the ISM by the destruction or death of stars.
\end{enumerate}

The terms $\kappa_{h}(x,y)$, $\kappa_{c}(x,y,t)$, $\kappa_{s}(x,y)$ and $\kappa_{a}(x,y)$ are
therefore, the proportionality factors of the SF in the halo, the cloud formation, the cloud-cloud collision, and the cloud-massive stars interactions (the last two create stars from molecular clouds), respectively. Since stars are divided into two groups, $s_{1}$, and $s_{2}$, the parameters involving star formation are divided into two groups, as well; thus: $\kappa_{h}=\kappa_{h,1}+\kappa_{h,2}$, $\kappa_{s}=\kappa_{s,1}+\kappa_{s,2}+\kappa_{s}' $, and $\kappa_{a}=\kappa_{a,1}+\kappa_{a,2}+\kappa_{a}'$, where  terms $\kappa_{s}'$ and $\kappa_{a}'$ refer to the restitution of diffuse gas due to cloud-cloud collisions and massive stars-cloud interaction processes, respectively.

In the following equations, we will omit the $x,y,t$ dependence of the already defined quantities to alleviate any potentially confusing notation. Thus, the star formation law in the halo and disc are defined:
\begin{eqnarray}
\Psi_{\textrm{\scriptsize H}}(x,y,t)& =& (\kappa_{h,1}+\kappa_{h,2})g_{\textrm{\scriptsize H}}^{n}\\
 \Psi_{\textrm{\scriptsize D}}(x,y,t) &=& (\kappa_{s,1}+\kappa_{s,2})c^{2}+(\kappa_{a,1}+\kappa_{a,2})c\,s_{\textrm{\scriptsize{2,D}}}
 \nonumber
\end{eqnarray}
,where $c$ is the molecular gas, which itself forms from diffuse gas and suffers from destruction processes (v) and (vii), leading to:

\begin{equation}
\frac{dc}{dt} =  \kappa_{c}\, g_{\textrm{\scriptsize D}}-\kappa_{s}\,c^2- \kappa_{a}\,c\,s_{\textrm{\scriptsize {2,D}}}. 
\end{equation}

The equations to calculate $\kappa_{h}$,  $\kappa_{s}$,  and $\kappa_{a}$ were given in \citet{fer94} and \citet{molla14}; each depends upon the volume of each halo or disc region, and are
derived via various {\sl efficiency} factors, $\epsilon_{h}$, $\epsilon_{s}$, and $\epsilon_{a}$. These efficiencies represent probabilities (in the range $[0,1]$) associated with processes of conversion among the different phases and are assumed to be constant within a galaxy. The efficiency to form stars in the halo, $\epsilon_{h}$, is obtained through the selection of the best value able to reproduce the SFR and abundances of the Galactic halo \citep{fer94} and takes a value between $1\times10^{-3}$ and $3\times10^{-2}$. The parameter $\kappa_{a}(x,y)=\kappa_{a}$, also appears without any dependence on the position of the region. The corresponding efficiency, $\epsilon_{a}$, which includes the interactions between clouds and massive stars --local processes-- was also obtained from the best value for the MWG.  Since in the present work, all our disc regions have the same area (1\,kpc$^{2}$), by assuming that the height of the disc, $h$, is always the same, $h\sim 200$\,pc, all volumes are similar and therefore the corresponding parameters have no dependence on galactocentric distance. Thus, the parameter $\kappa_{s}$ is also the same for all our cells. 

For $\kappa_{\rm c}$ we use the same prescription as in \citet{mol17}:
\begin{equation}
\kappa_{\rm c}(x,y,t) = 2.67\times \Sigma_{\rm gas} \, \left(\Sigma_{\rm gas} + \Sigma_{\star}\right) \times \left(Z + Z_{\rm ini}\right),
\end{equation}
where $Z=Z(x,y,t)$ is the total abundance of metals or metallicity, $Z_{\rm ini}=1.4\,10^{-5}$ is a threshold value for commencing star formation at early times, and $\Sigma_{*}$ and $\Sigma_{\rm gas}$ are the surface density for stars and total gas, respectively; that is, $\Sigma_{\rm gas}=\Sigma_{\rm g_{D}}+\Sigma_{\rm c}$, and the value 2.67 has units of Gyr$^{-1}\,{\rm M}_{\sun}^{-2}\,{\rm pc}^{2}$. 

\subsection{Model equations}
\label{eqsys}

Having defined the initial spatial distributions of mass, and the relevant physical processes implemented within our framework, it 
is necessary to solve the {\sc 2D-MulChem} system of differential equations that determines the evolution in time of each region located at $(x,y)$:
\begin{small}
\begin{eqnarray}
\nonumber
\frac{dg_{\textrm{\scriptsize H}}}{dt} & = & -\left(\kappa_{h,1}+\kappa_{h,2}\right)g^{n}_{\textrm{\scriptsize H}}-f\, g_{\textrm{\scriptsize H}} +W_{\textrm{\scriptsize H}} 
\nonumber \\
\frac{ds_{\textrm{\scriptsize {1,H}}}}{dt} & = & \kappa_{h,1}g^{n}_{\textrm{\scriptsize H}}-D_{\textrm{\scriptsize{1,H}}} \nonumber \\
\frac{ds_{\textrm{\scriptsize{2,H}}}}{dt} & = & \kappa_{h,2}g^{n}_{\textrm{\scriptsize H}}-D_{\textrm{\scriptsize{2,H}}} \nonumber \\
\frac{dg_{\textrm{\scriptsize D}}}{dt} & = & -\kappa_{c} g_{\textrm{\scriptsize D}}+\kappa_{a}'cs_{\textrm{\scriptsize{2,D}}} +\kappa_{s}'c^{2}+f\, g_{\textrm{\scriptsize H}}+W_{\textrm{\scriptsize D}}
\nonumber\\
\frac{dc}{dt}& =& \kappa_{c}g_{\textrm{\scriptsize D}}-\left(\kappa_{a,1}+\kappa_{a,2}+\kappa_{a}' \right)c \, s_{\textrm{\scriptsize{2,D}}} \nonumber\\
& & -\left(\kappa_{s,1}+\kappa_{s,2}+\kappa_{s}'\right)c^{2}
\nonumber \\
\frac{ds_{\textrm{\scriptsize{1,D}}}}{dt}& =& \kappa_{s,1}c^{2}+\kappa_{a,1}c \, s_{\textrm{\scriptsize{2,D}}}-D_{\textrm{\scriptsize{1,D}}}
 \nonumber \\
\frac{ds_{2,D}}{dt}& =& \kappa_{s,2}c^{2}+\kappa_{a,2}c \, s_{\textrm{\scriptsize{2,D}}} -D_{\textrm{\scriptsize{2,D}}}
\nonumber \\
\frac{dr_{\textrm{\scriptsize H}}}{dt} & =& D_{\textrm{\scriptsize{1,H}}}+D_{\textrm{\scriptsize{2,H}}}-W_{\textrm{\scriptsize H}}
\nonumber \\
\frac{dr_{\textrm{\scriptsize D}}}{dt} & = & D_{\textrm{\scriptsize{1,D}}}+D_{\textrm{\scriptsize{2,D}}}-W_{\textrm{\scriptsize D}},
\end{eqnarray}
\end{small}
where $D_{H,D}$ are the death rates, which depend on the average lifetime of stars of mass $m$, through the following:
\begin{equation}
D_{1;H,D}(x,y,t)=\int_{m_{min}}^{m_{*}}\Psi_{H,D}(x,y,t-\tau_{m})m\phi(m)dm
\end{equation}
\begin{equation}
D_{2;H,D}(x,y,t)=\int_{m_{*}}^{m_{max}}\Psi_{H,D}(x,y,t-\tau_{m})m\phi(m)dm
\nonumber
\end{equation}
The quantities $g$, $c$, and $s$ denote the masses of diffuse gas, molecular clouds, and stars, respectively, in each region $(x,y)$, and the sub-index $H$ or $D$ corresponds to the halo or the disc component.
We assume that stars are created in the range $m_{\rm min}=0.15\,M_{\sun}$ to $m_{\rm max}=100\,M_{\sun}$, dividing them into two types: low mass stars with $m<4\,{\rm M}_{\odot}$, which corresponds to sub-indices $1$, and intermediate mass and massive stars, $m\ge 4\,{\rm M}_{\sun}$, which correspond to sub-indices $2$.
Each equation defines the conversion of mass from one phase to the other, as explained in Section~\ref{sfr}. The terms $W_{H, D}$ are defined in the next subsection.

\subsection{Stellar yields and element ejecta}
\label{yields}

The equations governing the evolution of the chemical abundances are equivalent to those given in \citet{mol17}:
\begin{eqnarray}
\frac{dX_{i,H}}{dt}& = &\frac{W_{i,H}-X_{i,H} W_{H}}{g_{H}}\\
\frac{dX_{i,D}}{dt}&=&\frac{W_{i,D}-X_{i,D}W_{D}+f \, g_{H}[X_{i,H}-X_{i,D}]}{g_{D}+c}
\nonumber
\end{eqnarray}
where $X_{i} = X_{i}(x,y,t)$ are the mass fractions of the 15 isotopes considered by the model:
$^{1}$H, D, $^{3}$He, $^{4}$He, $^{12}$C, $^{16}$O, $^{14}$N, $^{13}$C, $^{20}$Ne, $^{24}$Mg, $^{28}$Si, $^{32}$S, $^{40}$Ca,
$^{56}$Fe, and the neutron-rich CNO isotopes. The restitution rates, $W_{i;H,D}$ are:
\begin{eqnarray}
\label{res}
W_{i;H,D}(x,y,t)& = &
\int_{m_{\rm min}}^{m_{\rm max}} \left( \sum_{j}{\tilde{Q}}_{ij}(m)X_{j}(x,y,t-\tau_{m})\right) \nonumber\\
& & \times \Psi_{H,D}(x,y,t-\tau_{m})dm,
\end{eqnarray}
which give the total production of each element $i$ created by the other elements $j$, integrated via the initial mass function and taking into account the SFR which occurred at each time $t-\tau(m)$.

To compute the elemental abundances, we use the technique based on the {\sl $Q$-matrix} formalism \citep{tal73,fer92}. Each element $(i,j)$ of a matrix, $Q_{i,j}$ gives the proportion of a star which was initially element $j$, ejected as $i$ when the star dies. Thus,
\begin{eqnarray} 
Q_{i,j}(m)& =& \frac{m_{i,j,exp}}{m_{j}}\\
Q_{i,j}(m)X_{j}& = & \frac{m_{i,j,exp}}{m}
\end{eqnarray}

 If we take into account the number of stars of each mass $m$, given by the initial mass function (IMF), $\phi(m)$, that eject this mass $m_{i,j}$, 
 we have:
 \begin{equation}
 \tilde{Q}_{i,j}(m)=Q_{i,j}m \phi(m),
 \end{equation}
 and, therefore, the term $\sum\limits_{j}{\tilde{Q}}_{i,j}(m)X_{j}$ in Eq.~\ref{res} represents the mass of an element $i$ ejected by all stars of initial mass $m$ by taking into account all possible production channels.  Stellar yields 
 have usually been computed  assuming only solar relative abundances among the different elements at a given Z. However, the relative abundances of elements are neither always solar nor constant along the evolutionary time. The use of $Q$ matrices allows us to take into account possible differences of chemical composition within a given Z, relaxing the hypothesis of solar proportions in the ejection, since each element $i$ relates with its own sources \citep[see][for details about the different components $Q_{i,j}$]{pcb98}. 

Following \citet{mol15}, we use the stellar yield sets from \citet{lim03} and \citet{chi04} for massive stars, together with yields from \citet{gav05,gav06} for low and intermediate mass stars, combined with the IMF from \citet{kro01}. For supernovae type Ia (SNe-Ia) we use the time-dependent rates given by \citet{rlp00}. The stellar yields for SNe-Ia are those of \citet{iwa99}. As shown in \citet{mol15} and \citet{mol19}, the observational data for the MWG are well-reproduced with our 1D-{\sc MULCHEM} model. 

\subsection{The spiral wave over-density}
\label{sec-sw}

\begin{figure}
\includegraphics[width=0.45\textwidth,angle=0]{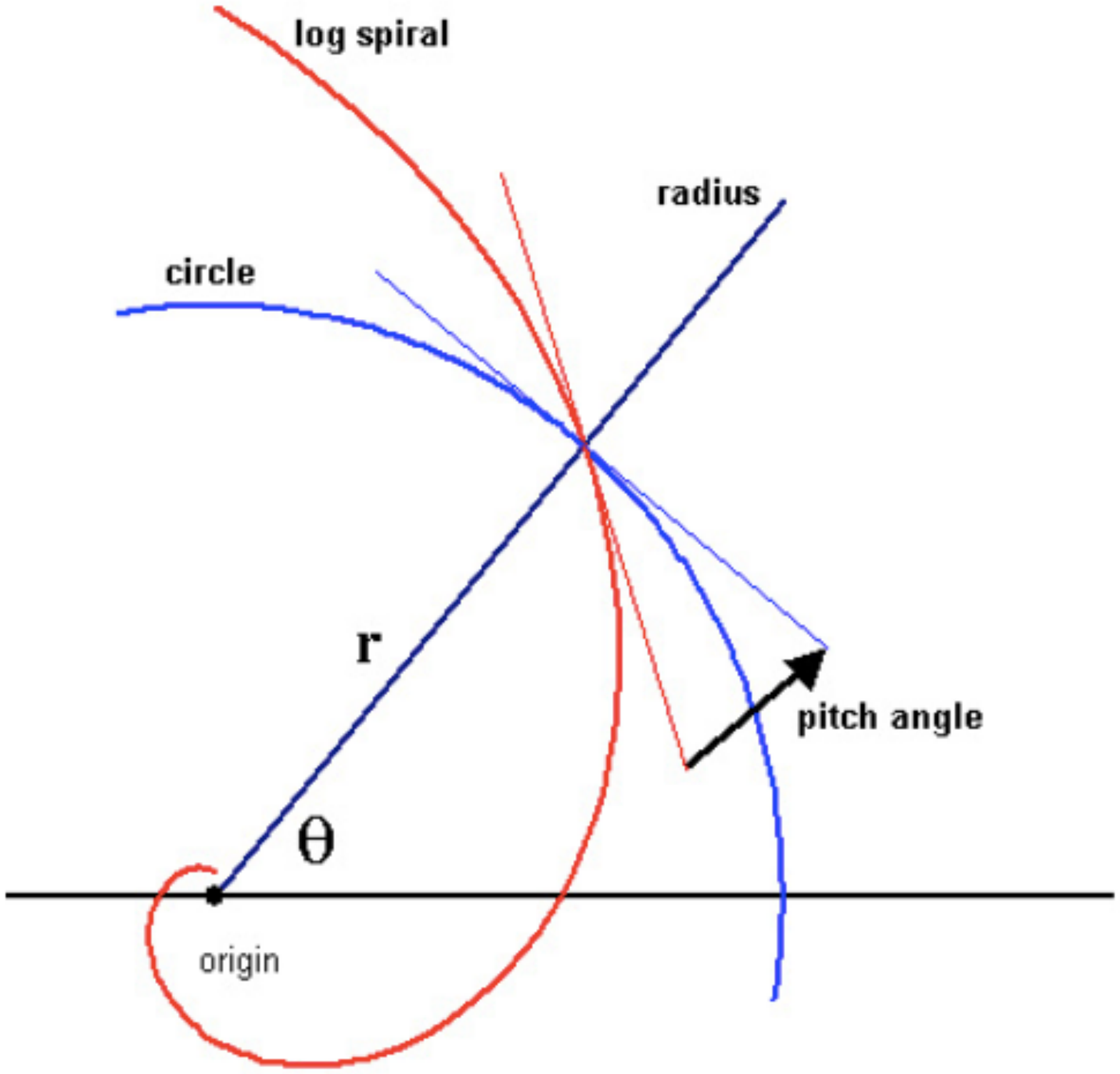}
\caption{Pitch angle $i$ description.} 
\label{pitch_angle}
\end{figure}
\label{swo}

 \begin{table}
 \caption{Properties of spiral arms}
 \centering
 \label{arm}
 \begin{tabular}{c c c c}
  \hline
  Property & symbol & value & unit  \\
\hline
Number of arms & m & 2 & - \\
Pitch angle & i & 14 & -  \\
Half width & $\sigma$ & 4.7 & kpc\\
Perturbation amplitude & $\zeta_{0}$ & 600 & $\textrm{km}^{2}$ $\textrm{s}^{-2}$ $\textrm{kpc}^{-1}$ \\
Scale length & $\varepsilon_{s}^{-1}$ & 2.5 & kpc \\
Spiral Wave crossing distance  & $R_{i}$ & 1.618 & kpc \\
Angular speed & $\Omega_{p}$ & 23 & km\,s$^{-1}$\,kpc$^{-1}$\\
\hline
 \end{tabular}
\end{table}

\begin{figure}
\begin{center}
\includegraphics[width=0.55\textwidth,angle=0]{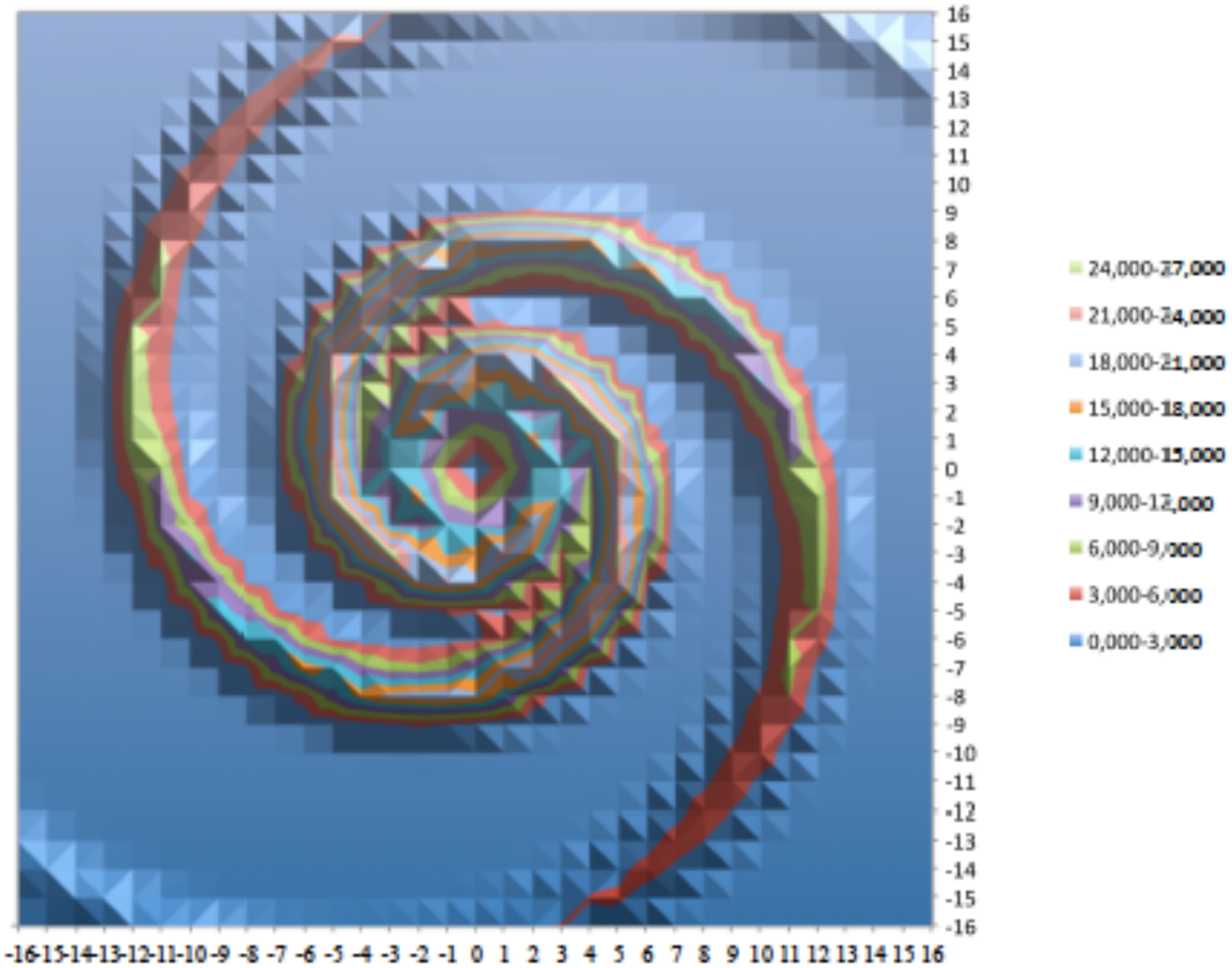}
\caption{Surface density for the spiral wave corresponding to equations from Section~ \ref{sec-sw} computed with parameters from Table 2. 
We include it at the initial time of the regions, $\Sigma_{SW}(x,y)$ on the disc mass for models SWD, SWS and SWR. This  corresponds to the halo in model SWH.See section\ref{modelos} for definition and description of computed models.}
\label{mass-2D}
\end{center}
\end{figure}
In order to include the effects of a spiral wave, we follow \citet[][ hereinafter JUN13]{jun13}. The surface density of a disc may be represented as the sum of an axi-symmetric surface density and a perturbed surface density $\Sigma_{sw}(R,\phi)$, which represents the spiral pattern in a frame that rotates at a angular speed $\Omega_{p}$. This means that to our radial distribution of disc mass, we now add the surface density of the spiral wave, $\Sigma_{sw}(R,\theta)$, obtained by solving the Poisson equation, as per JUN13:
\begin{equation}
 \Sigma_{sw}(R, \theta, t)
 = \Sigma_{so}\, e^{ -\frac{R^2}{\sigma^2} \left[ 1-cos(m\varphi(t)-f_{m}(R)) \right] },
 \label{sw}
\end{equation}
where $\varphi$ is the azimuthal coordinate at the rotation frame of the spiral wave, and therefore $\varphi(t) = \theta-\Omega_{p}t$, with $\theta$ being the angle in the inertial frame, and
\begin{equation}
\Sigma_{so}(R,\theta)=\frac{\zeta_{0}m}{2\pi G}\frac{R^{2}}{\sigma^{2}\,\mid {\tan{i}}\mid}e^{-\varepsilon_{s}R},
\end{equation}
while  $f_{m}(R)$ denotes the shape function given by
\begin{equation}
 f_{m}(R)=\frac{m}{\tan{(i)}}ln(R/R_{i})+ \gamma,
\end{equation}
where $\zeta_{0}$ is the perturbation amplitude, $\varepsilon_{s}^{-1}$
is the scale length of the spiral arm, $m$ is the number of arms, $i$ is the pitch angle, $R_{i}$ is the
point where the spiral crosses the coordinate $x = 0$, $\sigma$ is the width of the Gaussian profile in the galactocentric azimuthal direction, $k = \frac{m}{R \tan{(i)}}$, and $\gamma$ is a phase angle. Eq.~\ref{sw} is obtained with the assumption of a zero-thickness disc and that of tightly-wound spiral arms in the plane $z = 0$. Figure ~\ref{pitch_angle} shows the definition of the pitch angle of a spiral wave, and we may also see this spiral wave compared with the circle or annulus defined by a radius $R$ and what $R_{i}$ would be in this case.
 
The properties of arms which have been considered in this paper are taken from JUN13 and summarised in Table~\ref{arm}. With these characteristics, the arms create a density contrast that is around 15-20\% \citep{ant11}, and take the shape shown in Figure~\ref{mass-2D}. The hypothesis being tested is that this spiral wave 
 might modify the SFR and, consequently, the elemental abundances in these regions with higher densities of gas compared with the ones where the density maintains the same values.

\begin{table}
 \caption{Characteristics of computed model}
 \label{models}
 \centering
 \begin{tabular}{c c c c c}
   \hline
  Model & SW &  Place & rotation & color\\
    name  & incl. &          &              & \\
\hline
AZ & no & .. &  ..& cyan\\
SWH & yes & halo & no & green\\
SWD & yes & disc & no & purple\\
SWS & yes  & disc & yes, SW & red \\
SWR & yes & disc & yes, SW+disc & blue\\
\hline
 \end{tabular}
\end{table}

\subsection{Computed models}
\label{compmodels}

We have computed five models whose characteristics are given in Table~\ref{models}.
The first (referred to as AZ) is the {\sl classical} model with azimuthal symmetry that we use to compare with the remaining four models; these 
four all include the spiral wave (SW), implemented as described in \S~2.5. 
In models SWH (Spiral Wave in the Halo) and SWD (Spiral Wave in the disc), we assume that the spiral arms are always in the same place with respect to the stars, and the over-density co-rotates with the galactic disc as if  $\Omega_{p}=V(R)/R$.
Model SWS takes into account the motion of the spiral pattern (S), while model SWR also includes the rotation (R) of the galactic disc.

In SWH, the mass corresponding to the arm is added to the halo mass in each region at the initial time:
\begin{equation}
g_{\textrm{\scriptsize{H}}}(x,y,0)
= \Delta\,M_{\textrm{\scriptsize{H}}}(x,y) + M_{\textrm{\scriptsize{SW}}}(x,y,0)
\end{equation}
where the first term $\Delta\,M_{\textrm{\scriptsize{H}}}(x,y)$ corresponds to Eq.~\ref{mhalo}, while the second one, $M_{\textrm{\scriptsize{SW}}}(x,y,0) = \Sigma_{sw}(R,\theta)\,A$, is obtained by multiplying the surface density perturbation $\Sigma_{sw}(R,\theta)$ given by Eq.~\ref{sw} for $t=0$, by the area $A = 1$\,kpc$^{2}$ of each individual region in the model.
One must take into account that this over-density is added to the halo mass, and it needs to fall over the equatorial plane forming the disc.
Therefore, the contrast may, at the end of the simulation, be lower than the empirical value of 15-20\%. 

In SWD, the spiral wave over-density given by Eq.~\ref{sw}, also evaluated at $t=0$, is added directly to the disc as a perturbation $\Delta\,g_{\textrm{\scriptsize{SW}}}(x,y,t)$ that is assumed to grow proportional to the depletion of the halo gas (i.e. roughly proportional to the disc mass):
\begin{equation}
\Delta\,g_{\textrm{\scriptsize{SW}}}(x,y,t)=M_{\textrm{\scriptsize{SW}}}(x,y,0)\times\left(\frac{\Delta\,M_{\textrm{\scriptsize{H}}}(x,y,0)-g_{\textrm{\scriptsize{H}}}(x,y,t)}{\Delta\,M_{\textrm{\scriptsize{H}}}(x,y,0)}\right).
\label{prop_SW}
\end{equation}
Numerically, this is implemented as a term of the form $\frac{d\Delta\,g_{\textrm{\scriptsize{SW}}}}{dt}$ that is added to the variation of the diffuse gas in the disc $\frac{dg_{\textrm{\scriptsize D}}}{dt}$ at every time step.
Such a prescription ensures that the intended magnitude of the perturbation is reached at the end of the evolution, avoiding an unrealistically strong spiral pattern at early times.

In all our models, the galactic disc (gas and stars) is rotating according to the rotation curve $V(R)$ appropriate for a $M_{dyn}=10^{12}\,M_{\sun}$ system \citep{sal07}.
In fact, we have derived the radial distribution of the disc mass expected at the present time from this rotation curve, which yields the usual exponential profile.
In models SWH and SWD we have only added the spiral density wave as a perturbation over the original disc or halo density, assuming that both the disc and the wave rotate simultaneously.
We may consider that these two models, as well as AZ, are computed in the rotation reference frame.

In SWS, we develop a more realistic model by adding the corresponding mass over-density to the disc regions, as in SWD, but also including the rotation of the spiral arm.
In this case, as the spiral wave is also rotating, we have computed the surface density perturbation using Eq.~\ref{sw} with $\varphi(t) = \theta-\Omega_{p}\times\,t$.
As before in SWD, we also multiplied it by the halo depletion factor that approximately takes into account the growth of the disc:
\begin{equation}
\Delta\,g_{\textrm{\scriptsize{SW}}}(x,y,t)=M_{\textrm{\scriptsize{SW}}}(x,y,t)\times\left(\frac{\Delta\,M_{\textrm{\scriptsize{H}}}(x,y)-g_{\textrm{\scriptsize{H}}}(x,y,t)}{\Delta\,M_{\textrm{\scriptsize{H}}}(x,y)}\right).
\label{prop_SWS}
\end{equation}

Finally, we compute another model, SWR, where we take into account
both the rotation of the spiral wave and the rotation of the disc itself,so we use the difference $\Omega_{p}-\Omega$, instead of only $\Omega_{p}$,  
 to compute the angle $\varphi(t) = \theta-(\Omega_{p}-\Omega)\times\,t$.
Since the rotation velocity is almost flat for $R > 5$\,kpc, the angular velocity $\Omega(R)=V(R)/R$ of the disc is higher than $\Omega_{p}$ at the inner regions and lower at the outer ones, and the resulting $\Omega_{p}-\Omega$ is positive within the co-rotation radius (approximately coincident with the solar neighbourhood) and negative beyond.
The net effect of disc rotation is thus a reduction of the effective propagation speed of the spiral wave, varying with galactocentric distance.
The relevant equations are similar to those for SWS, with the exception of using $\Omega_{p}-\Omega$ for computing $\varphi(t)$.
It is expected that both models, SWS and SWR, mix the gas much more than the first two models without rotation.
Therefore, differences between arm and inter-arm regions should be smaller, without a clear dependence on the azimuthal angle. 

As explained, we compute a classical model with azimuthal symmetry using only the density distribution corresponding to Table~\ref{mwg-mass-R}. This is AZ, the basic model with which we compare the other four. This model, as implemented in 1D-{\sc MulChem} scheme, has been calibrated in \citet{mol15,mol17} with MWG data (radial distributions of diffuse and molecular gas, stars, and SFR, as well as elemental abundances of C, N, and O at the present time). Predicted abundance patterns at other times are described in \citet{mol19}.

\begin{figure*}
\centering
\includegraphics[width=1.1\textwidth,angle=0]{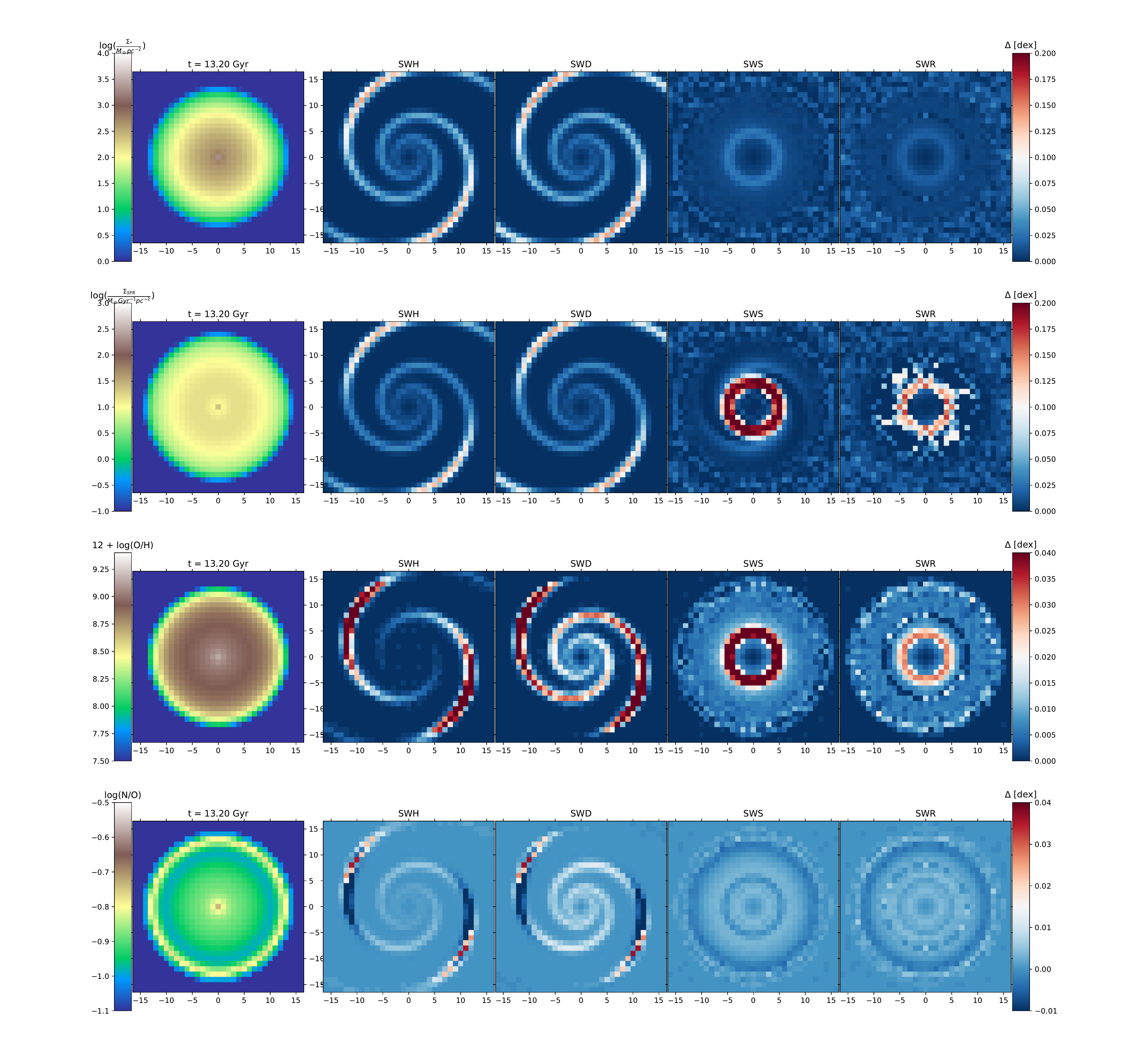}
\vspace{-1cm}
\caption{Results for AZ in the present time, column 1, and differences of our four  models, SWH, SWD, SWS and SWR, in columns 2, 3, 4 and 5, respectively, 
for the stellar surface density, first row, (in units of M\,$_{\sun}\,{\rm pc}^{-2}$), in logarithmic scale; the SFR surface density, 
(in units of M\,$_{\sun}\,{\rm pc}^{-2}\,{\rm Gyr}^{-1}$), second row,  ( in logarithmic scale as row one); 
for  oxygen abundance, as $12+\log{(O/H)}$, in third row; and for the relative abundance of nitrogen-to-oxygen ratio, as $\log(N/O)$ in the bottom.}
\label{diff-2D} 
\end{figure*}

\section{Results}
\label{results}

\subsection{Present time}

Here, we will compare the distributions of stellar mass surface density, SFR, and elemental abundances of O and N obtained from the AZ model at the present time with the results of the four other chemical evolution models, to see the effect of the spiral arm on these observables.

Bi-dimensional maps are plotted in Figure~\ref{diff-2D}, showing the results of the {\sl classical} AZ model in column 1, and the residuals of each model (SWH, SWD, SWS, and SWR) minus AZ in columns 2, 3, 4, and 5, respectively. The first row corresponds to stellar surface density, in logarithmic scale and $\rm M_{\sun}\,pc^{-1}$ units; the second one to the SFR surface density, also in logarithmic scale and  $\rm M_{\odot}\,pc^{-2}\,Gyr^{-1}$ units;  the third one to the oxygen abundances as $12+\log{(O/H)}$; and the final row to the nitrogen-over-oxygen relative abundance, written as $\log{(N/O)}$.
Since all are shown in decimal logarithm scale, differences between models with SW and AZ indicate the ratio between two quantities, in $dex$.

\begin{figure*}
\includegraphics[width=0.48\textwidth,angle=0]{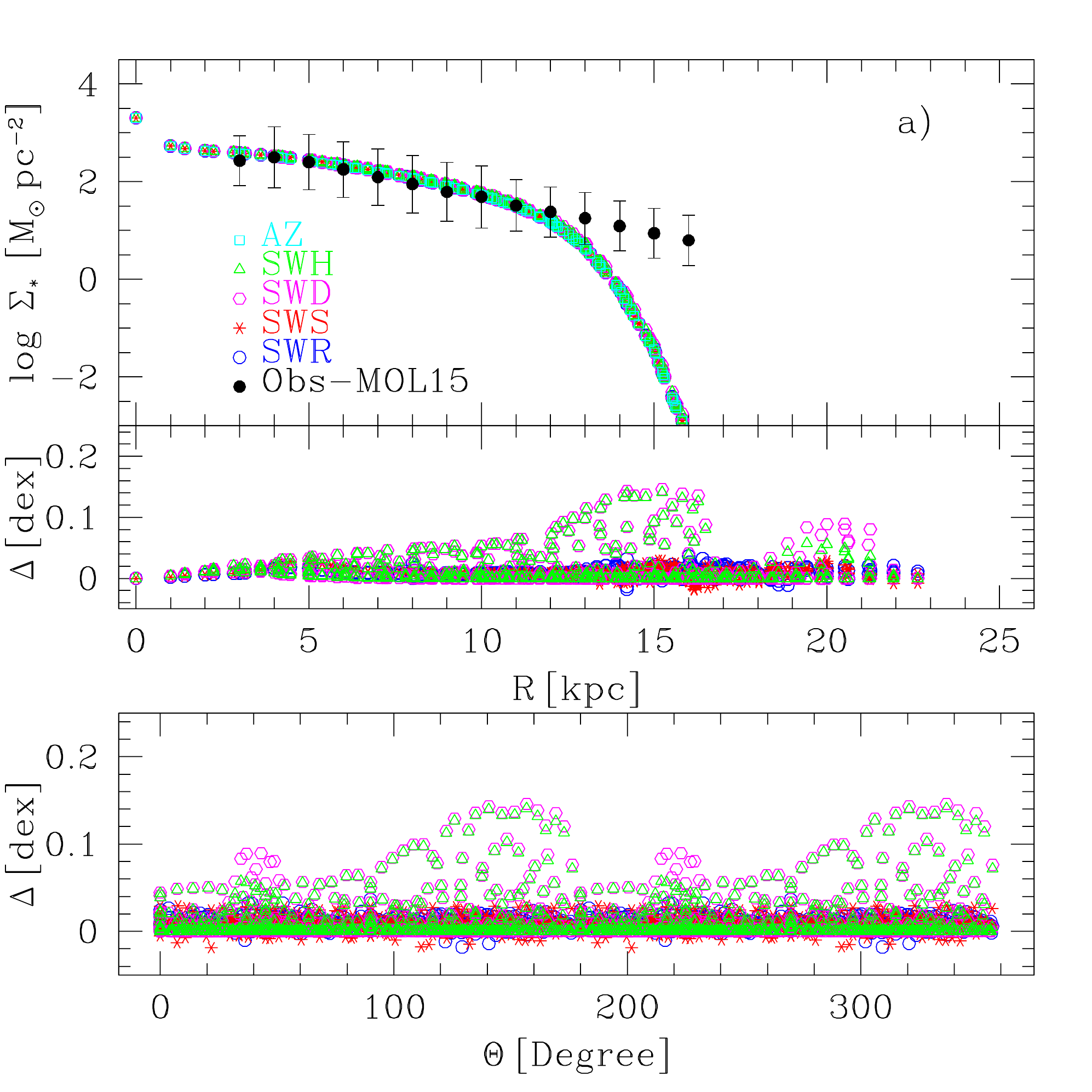}
\includegraphics[width=0.48\textwidth,angle=0]{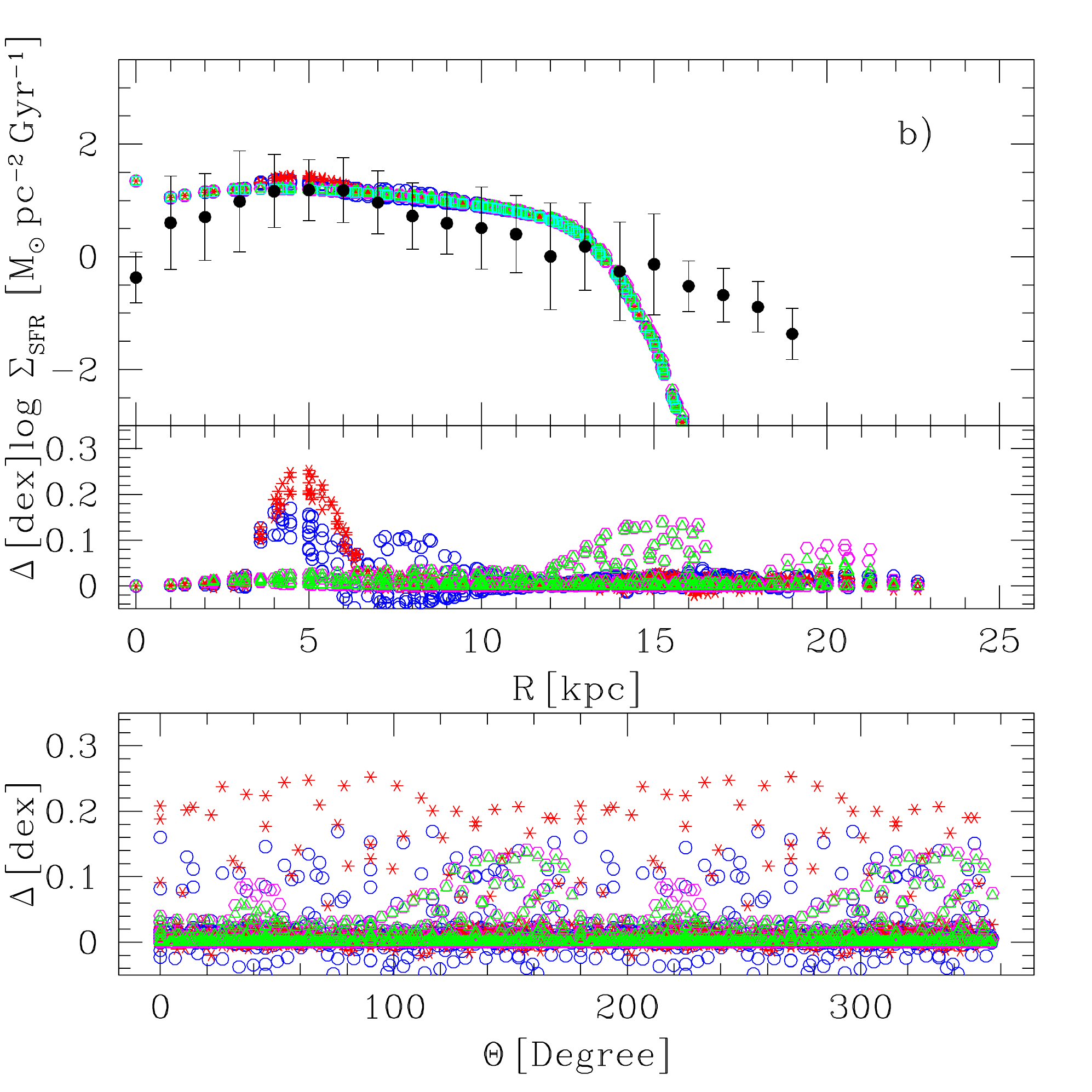}
\includegraphics[width=0.48\textwidth,angle=0]{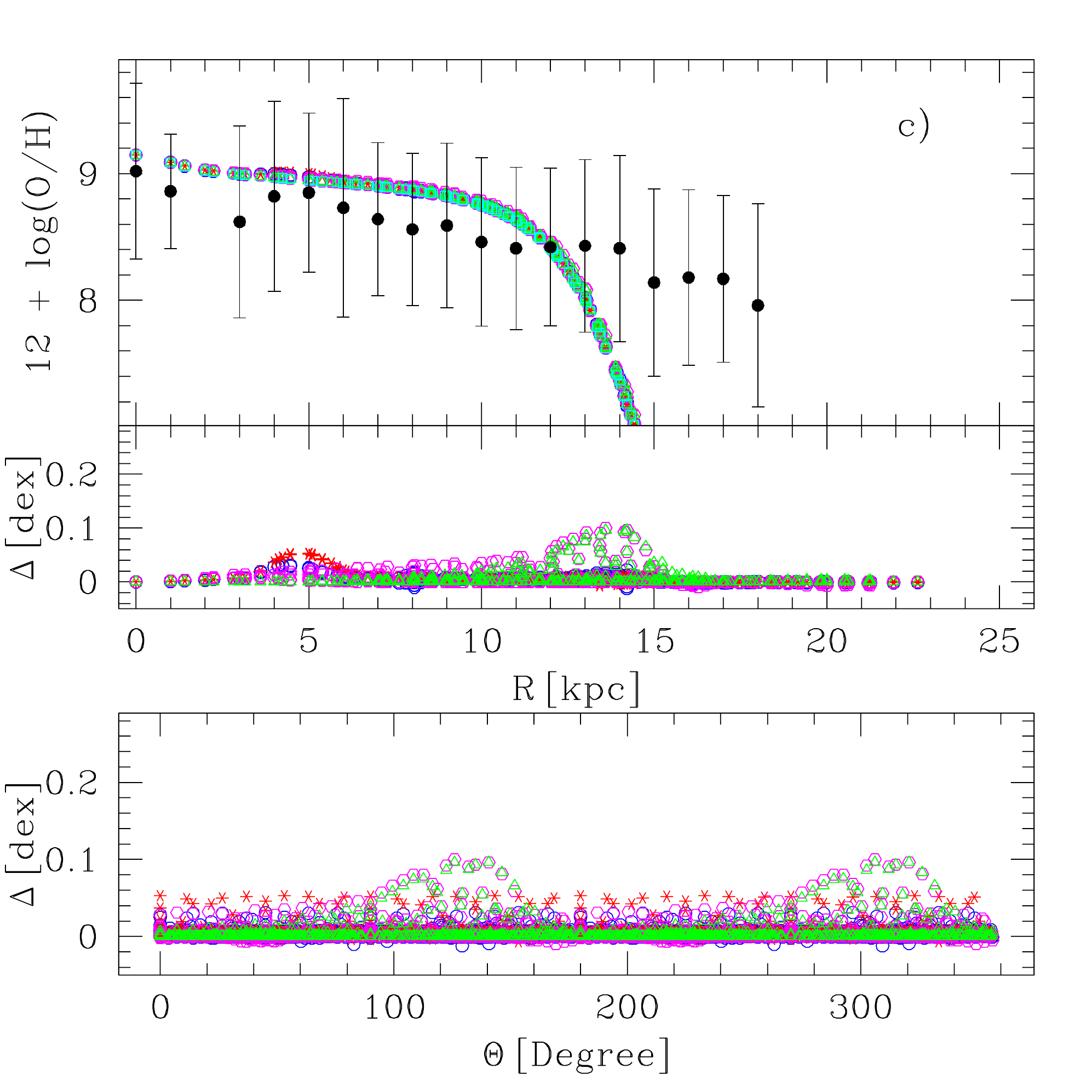}
\includegraphics[width=0.48\textwidth,angle=0]{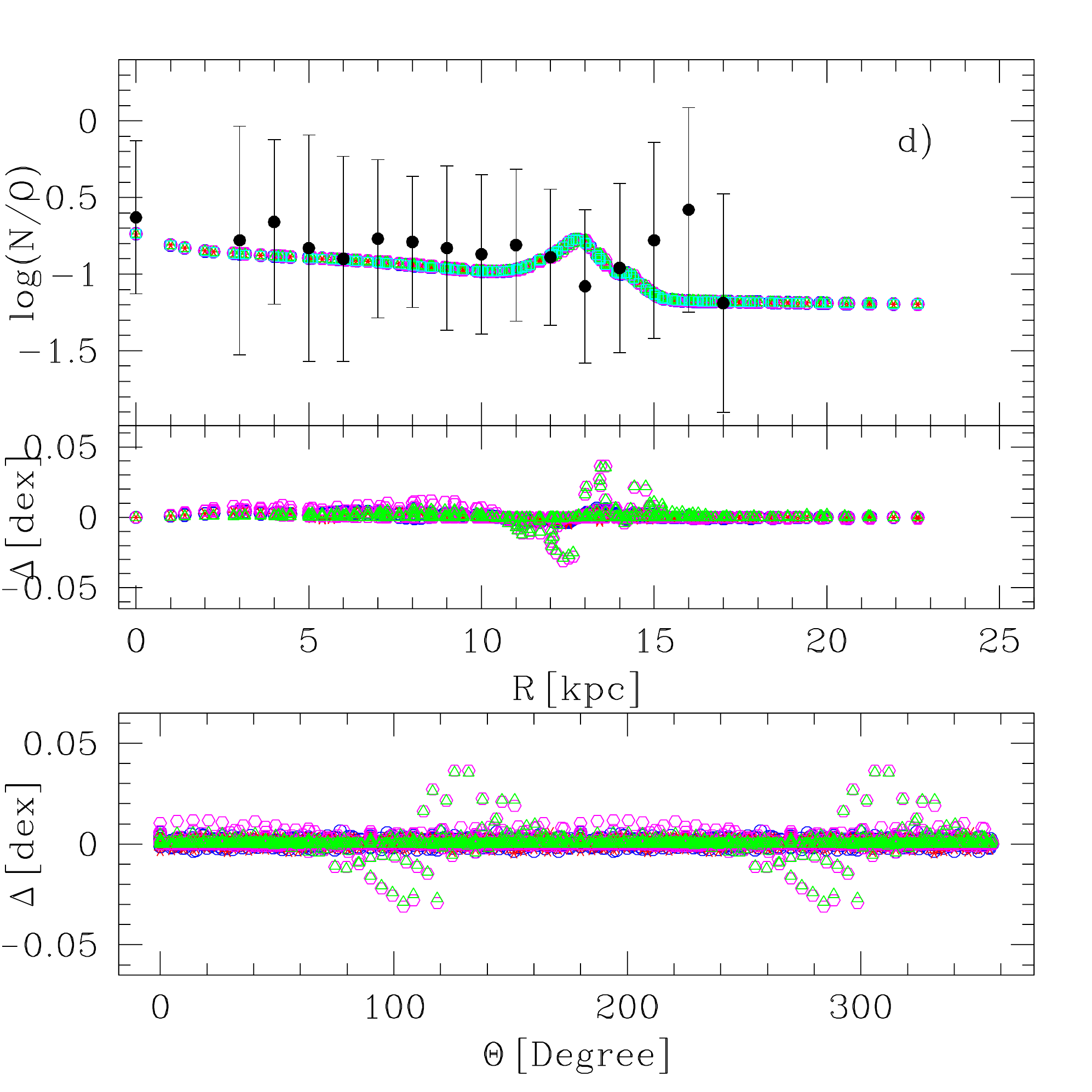}
\caption{The radial distribution for the present time of: a) stellar surface density; b) star formation rate surface density; c) elemental oxygen abundance $12+\log{(\textrm{O/H})}$; and d) for the relative abundance $\log(N/O)$, for all models as labelled, with the observational data in top panel of each. In the middle and bottom panels of each quantity, we show the differences between each model and AZ as a function of the radius and of the azimuthal angle $\theta$. }
\label{vs-R}
\end{figure*}

We also present the surface densities of stellar mass and star formation rate, the oxygen abundance and the nitrogen-to-oxygen ratio for the present time in Figure~\ref{vs-R}, as a function of the galactocentric radius $R$ and azimuthal angle $\theta$.
In addition to the differences between SW models and AZ, we also show absolute values in the top panels of Fig \ref{vs-R}a), b), c) and d), together with observational data compiled from the literature \citep{mol15}.

As shown in Figure~\ref{diff-2D}, the spiral wave manifests itself clearly in the present-day stellar surface density -- top row -- in models SWH  and SWD -- second and third columns, respectively.
Differences are of the order of $\sim 0.05-0.1$\,dex, consistent with our input of the spiral arm, with the highest values observed in the outer regions near the ends of the arms.
For the most realistic models, SWS and SWR, where the rotation of the spiral density wave (and the disc, in the latter) is implemented, the magnitude of stellar surface density variations compared with AZ are similar to SWH and SWD in the inner regions, but substantially smaller in the outer disk.
Most importantly, they do not show any spiral wave shape -- or any other azimuthal angle dependence, as shown in the bottom panel of Fig.~\ref{vs-R}a) -- due to the mixing effect of rotation.

Differences between each model and AZ for SFR are shown in the second row of Figure~\ref{diff-2D} as well as Figure~\ref{vs-R}b).
Once again, differences in SWH and SWD are small except in the outer regions of the disc ($R > 14$\,kpc).
There is also an azimuthal pattern with two maxima at 140 and 320 degrees, corresponding to the location of the spiral arms.
Models SWS and SWR reach fairly high values, up to $\sim 0.2$ dex, in a star-forming ring around 5\,kpc with a slight modulation pattern in the azimuthal angle.
We may still see the spiral wave in SWS at the present time but not in SWR, where it has been distorted by the disc rotation, even though a certain variation with $\theta$ is visible in the bottom panel of Fig.~\ref{vs-R}b) for these two last models, at variance with the stellar density profile.
Many of these effects should be detectable in H$\alpha$, where flux measurements with relative errors of the order of a few percent are routinely available.

In general, the effect of the spiral arms on the elemental abundances of O and N is smaller than in $\Sigma_{*}$ and SFR.
For $12+\log{\textrm(O/H)}$, shown in the third row of Figure~\ref{diff-2D} and panel c) of Figure~\ref{vs-R}, differences are always positive.
The maximum residual is $\sim 0.1$\,dex, almost half the value of that of the stellar or the SFR surface densities, and smaller or of the same order as the typical uncertainties in observational estimates of the oxygen abundance.
Still, the spiral wave is seen in both SWD and SWH, whereas the rotating models SWS and SWR do not show any spiral pattern.
The over-density due to the spiral wave hardly changes the distribution of the oxygen abundances because the mixing due to the rotation of the wave creates high abundances across the entire disc, homogenising the ISM and its abundances.
The abundance differences (always $> 0$) are more clearly seen as depending on the azimuthal angle in the non-rotating models.

The resulting simulations are here compared with the data compiled in our previous paper \citep{mol15}, and we see that they differ from the observational ones in the outer regions of the disc. However, the O abundances from the observational data were compiled from different sources (PNe, Cepheids, OB stars, HII regions) and authors and, therefore, small variations of the radial O gradient along the Galactic disc may be erased by this method. Indeed, there are some authors claiming that the O radial gradient cannot be fitted by a single-slope gradient, and it is necessary to take into account a possible bimodal distribution of the O gradient with different slopes in each region \citep[see e.g.][ and references therein]{maciel19, korotin14}. However, many authors claim that the distribution is actually flatter for R$> 13-15$\,kpc, at variance with our steeper one.

The nitrogen-over-oxygen abundance ratio, $\log{(N/O)}$, is shown in
the fourth row of Figure~\ref{diff-2D} and panel d) of
Figure~\ref{vs-R}. For N/O, differences are essentially non-existent
in SWS and SWR. In SWH and SWD, the N/O differences are even smaller,
even with negative residuals in some regions. We notices that N/O shows an abrupt change from positive to negative just before and after the spiral arm. This only occurs in the outer regions of the disc (mostly for $R > 16$\,kpc). In the rest of the disc, the spiral wave does not show a significant effect at the present time.
In fact, as averaged values, we can say that variations with azimuthal
angle are fairly negligible, and certainly smaller than the
observational error bars achievable at the present time. 

In all SWR panels, the co-rotation ring 8\,kpc is very clear,
corresponding to where both frequencies are equal:
$\Omega_{p}=\Omega(R)$, thus suppressing the time dependence of the
over-density in this model, doing results, mainly abundances, similar
to the AZ model along regions in this radius. However, the effect of
the spiral wave is apparent inside this co-rotation ring, creating a
radial distribution with a lobe. 

\cite{spi19} results are in agreement with most of these findings, the
largest fluctuations with azimuth being more evident in the outer
regions of the disk. However, with a spiral arm, they obtain
the largest variations in the corotation region, at variance with this
work, although our residuals are in excellent agreement with the small
values ( Res$_{OH} \le 0.01$\,dex in their model S2A) of the other
radii produced by these authors.

\subsection{Time evolution}
\label{time}

\begin{figure}
\centering
\includegraphics[width=0.45\textwidth,angle=0]{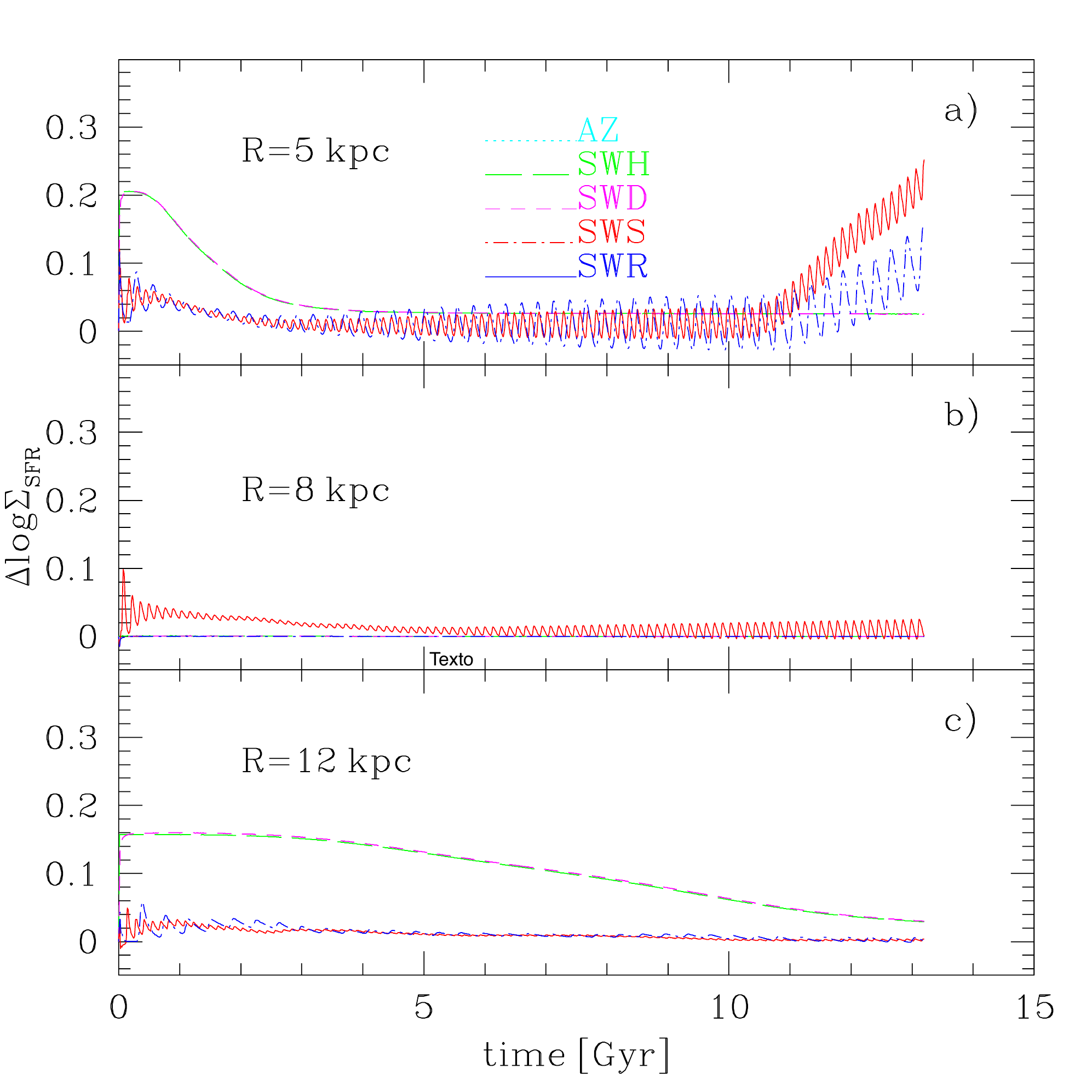}
\caption{Evolution with time of the SFR surface density as $\log{\Sigma_{\rm SFR}}$ in M$_{\sun}\,{\rm pc}^{-2}\,{\rm Gyr}^{-1}$ for three different regions located at galactocentric distances: a) $R=5$\,kpc; b) $R=8$\,kpc; and c) $R=13$\,kpc. Each model is shown as a different colour and line as labelled in panel b).}
\label{sfr_t}
\end{figure}

The results discussed in the previous section correspond to the state of the Galaxy at the present time. Since some authors claim that the arm tends to disappear with on Gyr timescales, or even shorter \citep{baba15}, we need to check at which moment of the evolution it will be necessary to add the arm to see its effects upon abundances. It would probably be better to include the arm at some other later time, such as $t \sim 12$\,Gyr, instead of $t=0$ (as is currently done). Although we have included the spiral wave and its corresponding over-density from the initial time $t=0$, we also have the results for the evolution from this time until the present, and so, we may learn something about how changes the effect of the spiral arm with time analysing the complete set of results and not only the final state described by the present time. The SFR must change as a consequence of the spiral density wave as it increases the density of gas in each cell and, if it changes with time due to the rotation, this will also produce a corresponding decrease. These expectations may be clearly seen in Figure~\ref{sfr_t}, where the evolution with time for the differences with AZ in the SFR surface density is represented for three cells, (5,0), (8,0) and (13,0) from top to bottom. Each model is shown by a different colour and line-type, as noted in panel a) and Table ~\ref{models}. There, we see the effect of the over-density on the SFR of different models. The cycles of increase and decrease suffered as a consequence of a rotating spiral wave (SWS and SWR models), with rapid variations due to the evolution of the wave with time, are clear. 

Model SWR shows smaller differences with AZ than SWS, due to the smaller final $\Omega_{p}-\Omega$ value used in the evolution of the spiral density wave. Moreover, in the co-rotation ring, the radius in which $\Omega_{p}=V/R$, the time evolution of the over-density disappears, similar to that seen in the same zone of model SWD.

It is expected, from dynamical studies, that the spiral arm had a stronger effect at the beginning of its evolution, and then, due to the mixing of stars and gas, differences tend to disappear with time. 
We have analysed this possibility by measuring the differences or residuals between each one of four models with a spiral wave and the AZ model, for the four quantities we have studied for the present time; therefore, we have calculated the mean values of these residuals with time, and their dispersion around these mean values:
\begin{equation}
<Res_{J}>(t)=\frac{\sum_{i}^{i=NT}(J_{SW}(t)-J_{AZ}(t))}{N}
\end{equation}
where J is each quantity, $\log{\Sigma_{*}}$, $\log{SFR}$, $12+\log{(O/H)}$ and $\log{(N/O)}$. NT is the total number of regions or spaxels calculated, 1089 ($33\times 33$) in our case, and SW indicates each one of the four models SWH, SWD, SWS and SWR.
\begin{figure*}
\includegraphics[width=0.45\textwidth,angle=0]{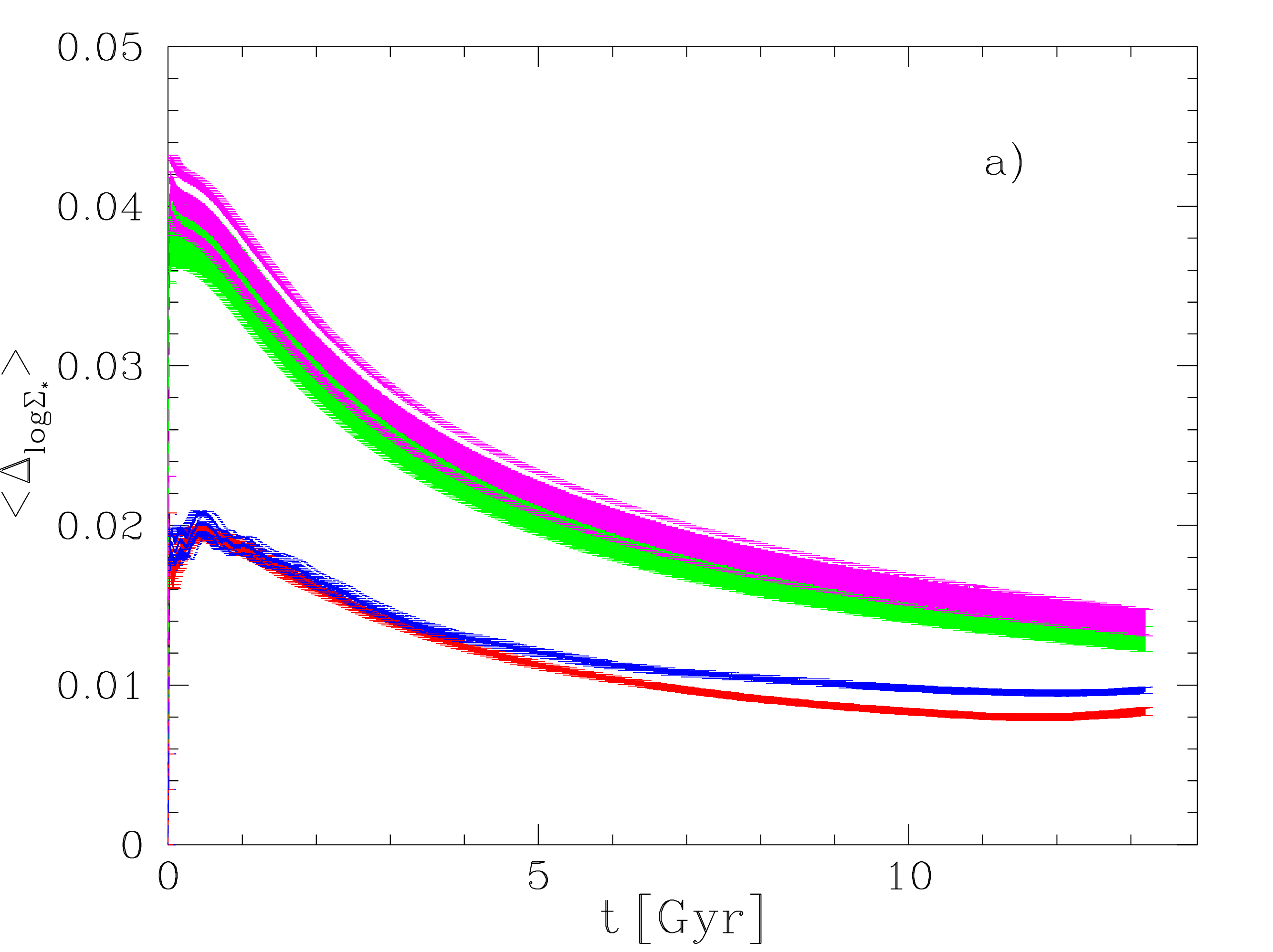}
\includegraphics[width=0.45\textwidth,angle=0]{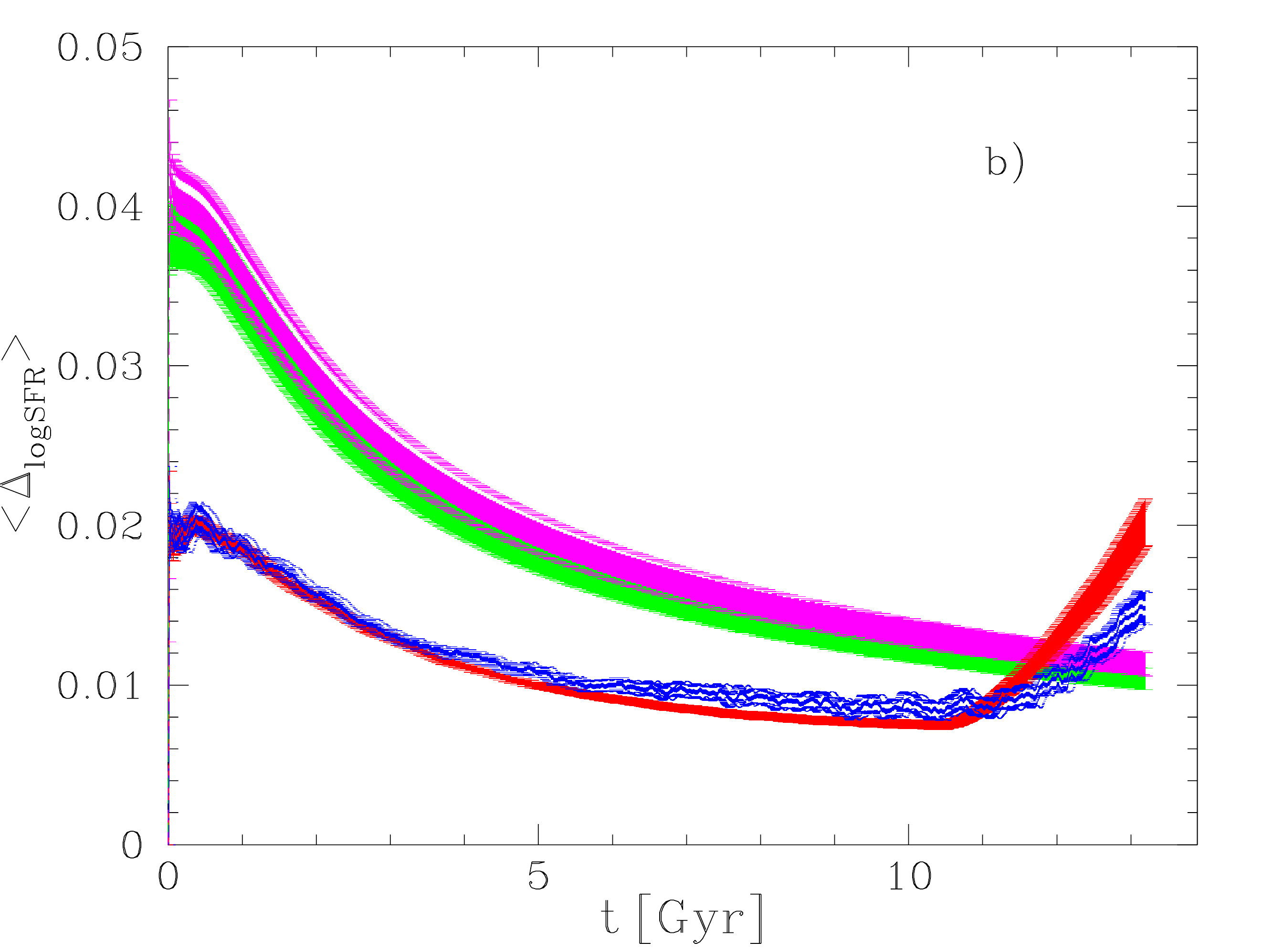}
\includegraphics[width=0.45\textwidth,angle=0]{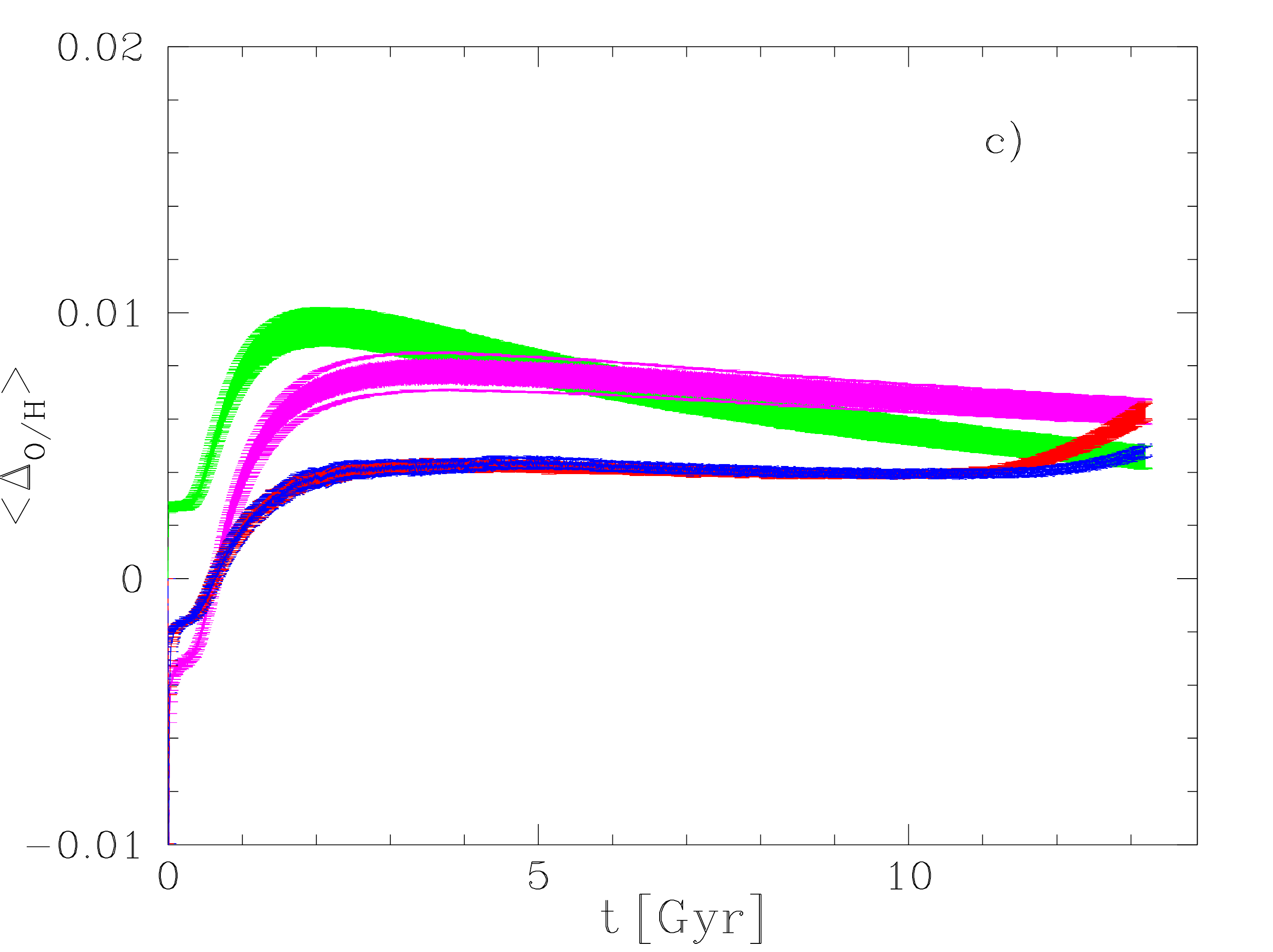}
\includegraphics[width=0.45\textwidth,angle=0]{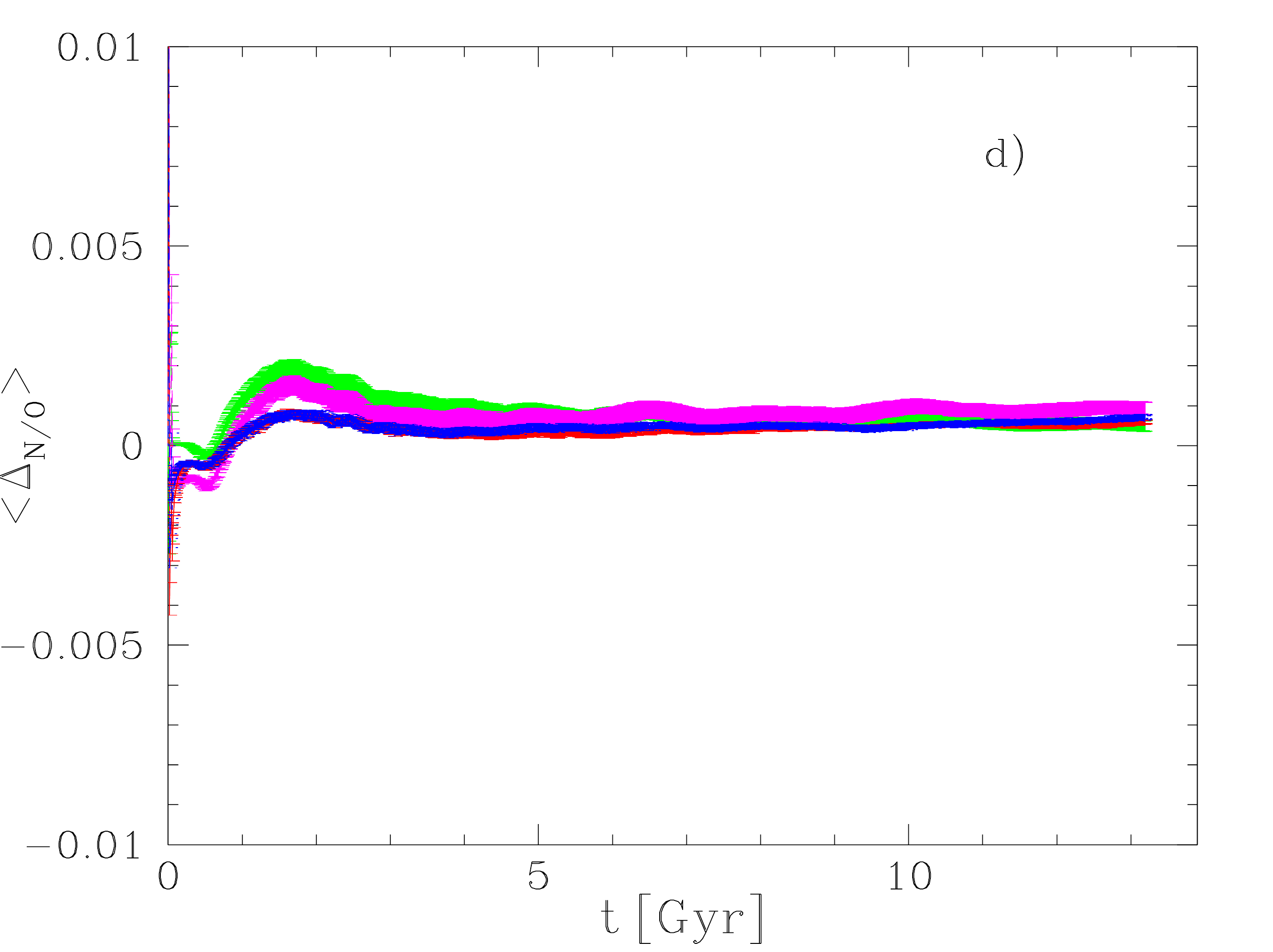}
\caption{Evolution along time of the mean residuals obtained for each model for: a) logarithm of stellar densities, $\log{\Sigma_{*}}$; b) logarithm of the star formation rate, $\log{\rm SFR}$; c) Oxygen abundances, $12+\log{(O/H)}$; and d) the relative abundance nitrogen-over-oxygen, $\log{(N/O)}$. Each model is shown as a different colour as in previous Figure~\ref{sfr_t}.}
\label{disp}
\end{figure*}

The corresponding dispersion, $\sigma$, comes from:
\begin{equation}
\sigma^{2}(t)=\frac{\sum_{i}^{i=NT}(Res_{i}-<Res>)^{2}}{NT\times(NT-1)}
\end{equation}

We show in Figure~\ref{disp} the evolution with time of these averaged residuals with their dispersion for: a) the logarithm of the stellar surface density,  $\log{\Sigma_{*}}$; b) the logarithm of the SFR, $\log{\rm SFR}$; c) the oxygen abundance, $12+\log{(O/H)}$; and d) the relative abundance ratio of nitrogen-to-oxygen, $\log{(N/O)}$.

For N/O, the differences are effectively constant and show very small variations with time in the four models. For the other three panels, however, differences with AZ decrease with time with maximum values occurring at $t \sim 1$\,Gyr  for stellar and SFR densities, and at  $t \sim 2$\,Gyr  for O/H and N/O. For SWS and SWR, the dispersion increases slightly at late times for SFR and O/H. The average with time and the maximum values of $<Res_{J}>$  corresponding to the present time,  are included in Table~\ref{residuals} with the corresponding dispersion. 

\begin{table*}[htb]
 \caption{Average residuals of SW models with time}
 \label{residuals}
 \begin{tabular}{ccccc}
 \hline
  Model & $Res_{\log \Sigma_{*}}$ & $Res_{\log \Sigma_{\rm SFR}}$ & $Res_{12+\log(O/H)}$ &  $Res_{\log(N/O)}$  \\
  \hline
  & \multicolumn{4}{c}{Mean Values}  \\
\hline
SWH &   0.0208 $\pm$   0.0002 & 0.0185 $\pm$ 0.0002 &  0.0068 $\pm$  0.0001 &  0.0008 $\pm$  0.0001\\
SWD &   0.0220 $\pm$   0.0002 & 0.0197 $\pm$ 0.0002 &  0.0064 $\pm$  0.0001 &  0.0008 $\pm$  0.0001\\
SWS &    0.0113 $\pm$   0.0001 & 0.0116 $\pm$ 0.0001 &  0.0038 $\pm$  0.0001 &  0.0004 $\pm$  0.0001\\
SWR &   0.0123 $\pm$   0.0001 & 0.0120 $\pm$ 0.0001 &  0.0040 $\pm$  0.0001 &  0.0005 $\pm$  0.0001\\
\hline
& \multicolumn{4}{c}{Present Time Values}  \\
\hline
SWH &    0.0130 $\pm$   0.0008 & 0.0104 $\pm$ 0.0007 &  0.0045 $\pm$  0.0005 &  0.0005 $\pm$  0.0002\\
SWD &    0.0140 $\pm$   0.0008 & 0.0113 $\pm$ 0.0007 &  0.0063 $\pm$  0.0005 &  0.0010 $\pm$  0.0002\\
SWS &     0.0090 $\pm$   0.0003 & 0.0200 $\pm$ 0.0015 &  0.0063 $\pm$  0.0003 &  0.0006 $\pm$  0.0001\\
SWR &    0.0097 $\pm$   0.0002 & 0.0148 $\pm$ 0.0010 &  0.0048 $\pm$  0.0002 &  0.0007 $\pm$  0.0001\\
\hline
 \end{tabular}
\end{table*}

Interestingly, and as we expected, the SWR model displays smaller averaged residuals than SWS, except for the stellar density. In fact, the averaged residuals take values in the range $\sim 0.01--0.04$\,dex for the stellar density, and  SFR, in SWH and SWD but only reach to 0.02 in SWR and SWS. These residuals are smaller for O/H and N/O in all models and for all times. These residuals are difficult to observe since they do not show an azimuthal dependence. 

This dependence could, however, be seen at the early times of the evolution of the spiral wave, before they disappear. It must be noted, though, that the maximum must occur just when the spiral density wave is produced. In that moment any perturbation may produce a Fourier wave which creates the density waves, until the system relaxes. Then, the spiral density wave disappears. However, the initial cause or perturbation may occur again, and in this way the spiral density wave would be a recurrent process. 
The arm induces strong kinematic imprints \citep{ant11}, and therefore a more sophisticated model, well beyond the scope of this work, would be necessary to properly take into account the time evolution of the spiral wave.

\begin{figure*}
\centering
\includegraphics[width=1.1\textwidth,angle=0]{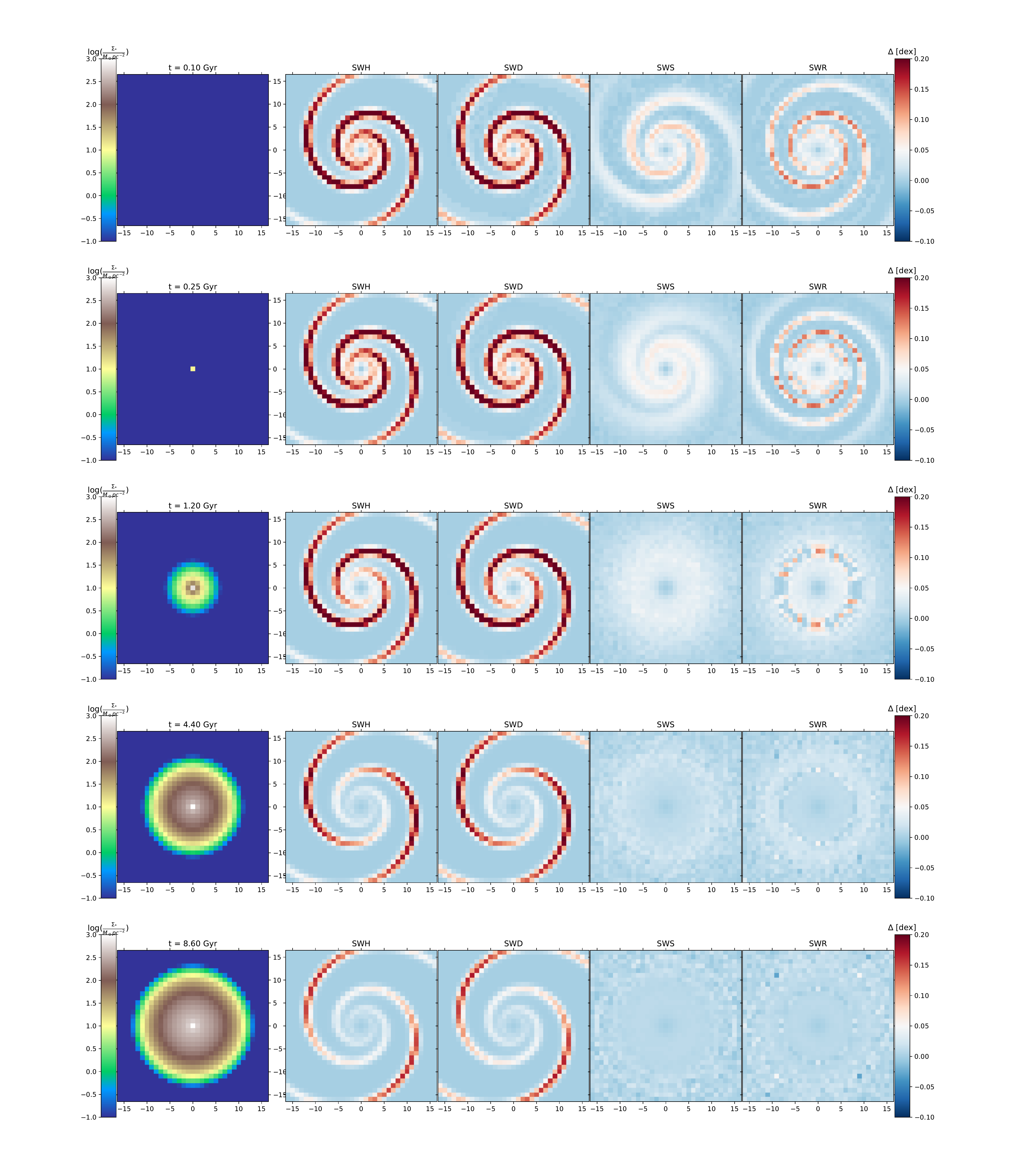}
\vspace{-1cm}
\caption{Time evolution of the stellar surface density  $\Sigma_{*}(x,y)$ in logarithmic scale at: 1)) $t=0.1$\,Gyr; 2) $ t=0.25$\,Gyr; 3) $t=1.20$\,Gyr; 4) $ t=4.40$\,Gyr; and bottom)  $t=8.60$\,Gyr. 
Left panels correspond to the AZ model, with the scale at the left side; 
while the four right columns are the differences of SW's models compared with AZ results: second column SWH-AZ; third column, SWD-AZ, fourth column SWS-AZ and the fifth, SWR-AZ. 
In these four cases we represent differences with the same color code: from blue (negative differences) to red (the positive ones).}
\label{sigma_est_t_2D}
\end{figure*}

Bearing such caveats in mind, let us now plot, in a similar way to Figure~\ref{diff-2D}, the differences with respect to the AZ model for different times.
We show in Figure~\ref{sigma_est_t_2D} the stellar surface density for AZ in column 1 and, as before, the differences with respect to it, SWH-AZ, SWD-AZ, SWS-AZ and SWR-AZ, in columns 2, 3, 4, and 5, respectively. Rows 1 to 5 correspond to times 0.10, 0.25, 1.20, 4.40 and 8.60\,Gyr.
\begin{figure*}
\includegraphics[width=1.1\textwidth,angle=0]{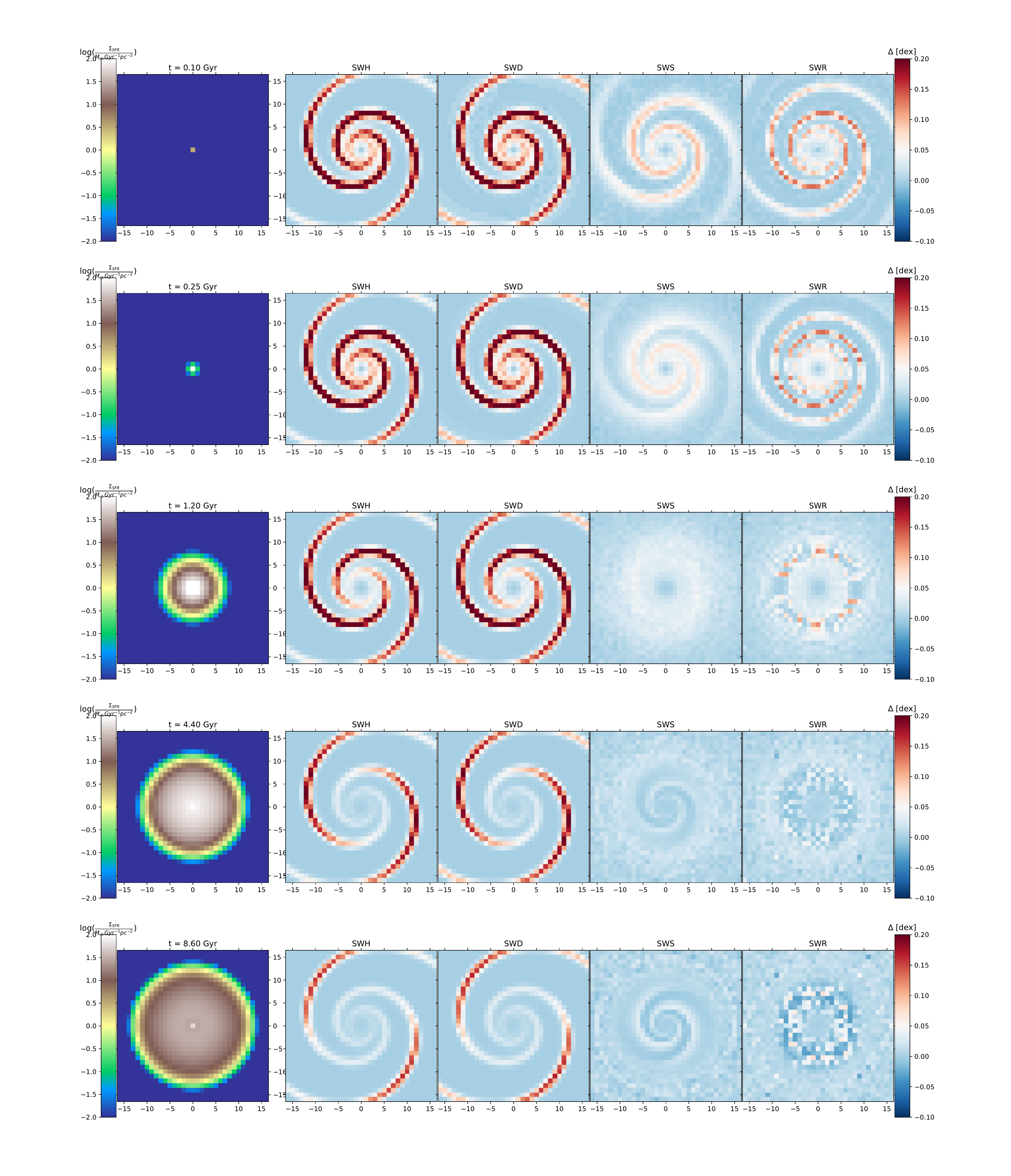}
\vspace{-1cm}
\caption{Time evolution of the SFR surface density  $\Sigma_{\rm SFR}(x,y)$ in logarithmic scale at: top)  $t=0.10$\,Gyr; second row) $t=0.25$\,Gyr; third row) $t=1.20$\,Gyr; fourth row) $t=4.40$\,Gyr; and bottom) $t=8.60$\,Gyr. Left panels correspond to the AZ model, with scale at the left side; while the right columns are the differences of SW's models compared with AZ results: second column SWH-AZ; third column, SWD-AZ, fourth column SWS-AZ and fifth column SWR-AZ.  In these cases we represent differences with the same color code: from blue (negative differences) to red (the positive ones).}
\label{sigma_sfr_t_2D}
\end{figure*}

It is evident in this figure that differences with AZ in SWS and SWR are similar, although slightly smaller than those from SWH and SWD, showing the spiral shape, in the first two/three times. After this, however, the disc homogenizes, and trends along the azimuthal angle tend to disappear. For SWH and SWD, however, differences are always small, showing values smaller than or similar to the uncertainties of the observational data, but showing a spiral wave shape decreasing along time.

In Figure~\ref{sigma_sfr_t_2D} we represent in a similar way, the corresponding SFR surface density. The figure has again five columns for AZ and SWH-AZ, SWD-AZ, SWS-AZ and SWR-AZ, as for Fig.~\ref{sigma_est_t_2D}, with the same times in rows 1 to 5. As in the stellar density distribution, the residuals are similar for all models in the two first times, $t=0.1$ and 0.25\,Gyr, but they dilute in SWS and SWR afterwards, showing values smaller than 0.05\,dex from t=4\,Gyr onwards. These differences are also small in SWH and SWD except in the very outer regions of the arms.
The differences, however, are higher for the final times for SWS than for SWR, showing more clearly a spiral wave shape in SWS.  This implies a different time evolution in the model SWR and a stronger mixing, while SWS still maintains its wave shape. 
 
\begin{figure*}
\hspace{-1cm}
\includegraphics[width=1.1\textwidth,angle=0]{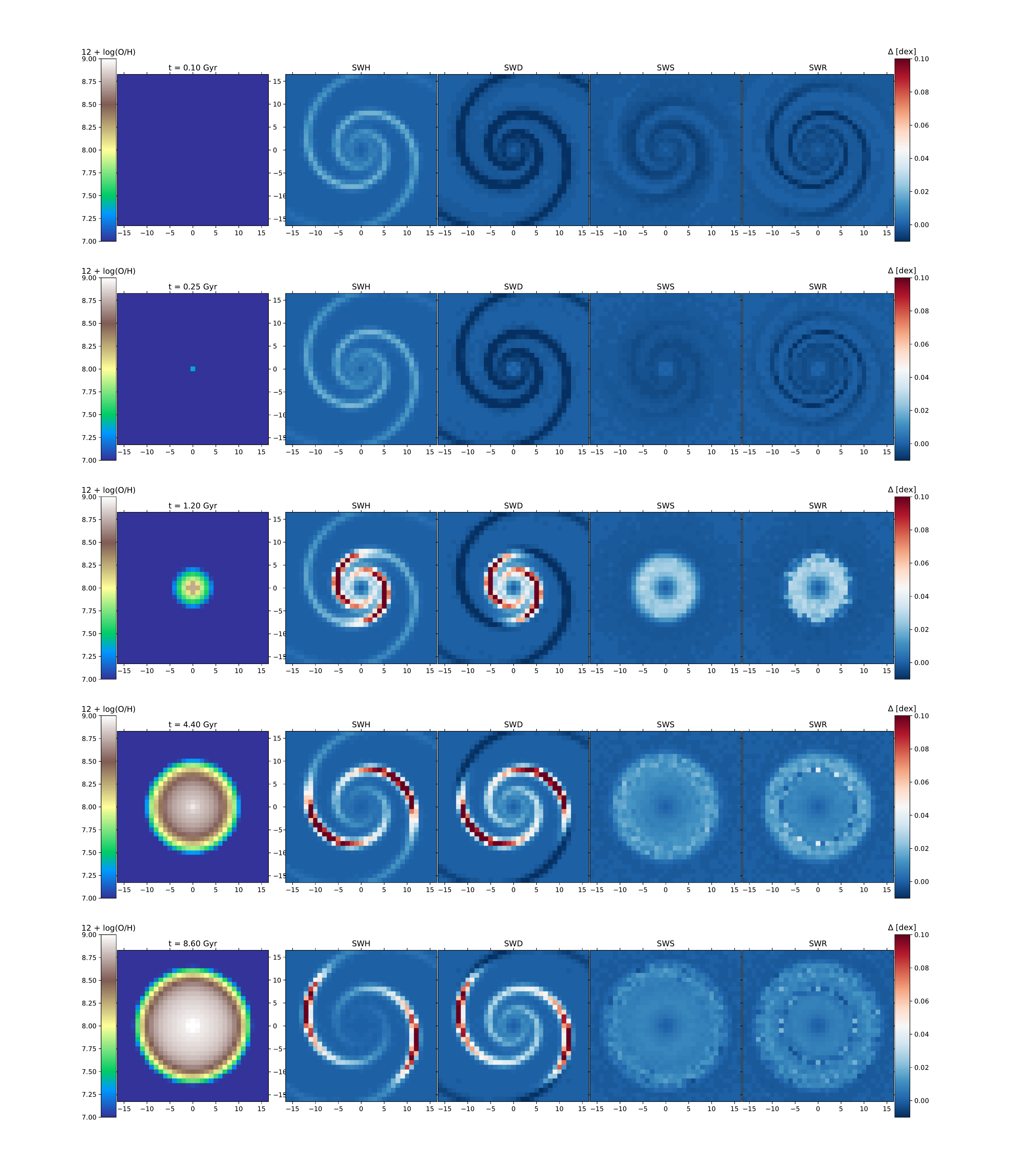}
\vspace{-1cm}
\caption{Time evolution of the oxygen abundance $12+\log{(O/H)}(x,y)$  at:   top)  $t=0.1$\,Gyr; second row) $t=0.25$\,Gyr; third row) $t=1.20$\,Gyr; fourth row) $t=4.40$\,Gyr; and bottom) $t=8.60$\,Gyr. Left panels correspond to the AZ model, with scale at the left side; while the right columns are the differences of SW's models compared with AZ results as in the previous Fig.~\ref{sigma_sfr_t_2D}.}
\label{OH_t_2D}
\end{figure*}

In Figure~\ref{OH_t_2D}, we show the time evolution of oxygen abundance, $12+\log{(O/H)}$, as per Figures~\ref{sigma_est_t_2D} and ~\ref{sigma_sfr_t_2D}.
Model SWH shows little differences with AZ at the first three times, until $t\le 1.2$\,Gyr, with residuals being below 0.05\,dex for later times except in the outer regions of the disc.
SWD displays smaller differences than SWH in the first two rows, but they are still observed in the last times shown.
Models SWS and SWR show small residuals after 4 Gyr, with the spiral
arm being visible only before this time. However, for the later times,
differences for all models are equally erased as a function of the
angle. For SWR, the co-rotation ring is very clear.

\begin{figure*}
\hspace{-1cm}
\includegraphics[width=1.1\textwidth,angle=0]{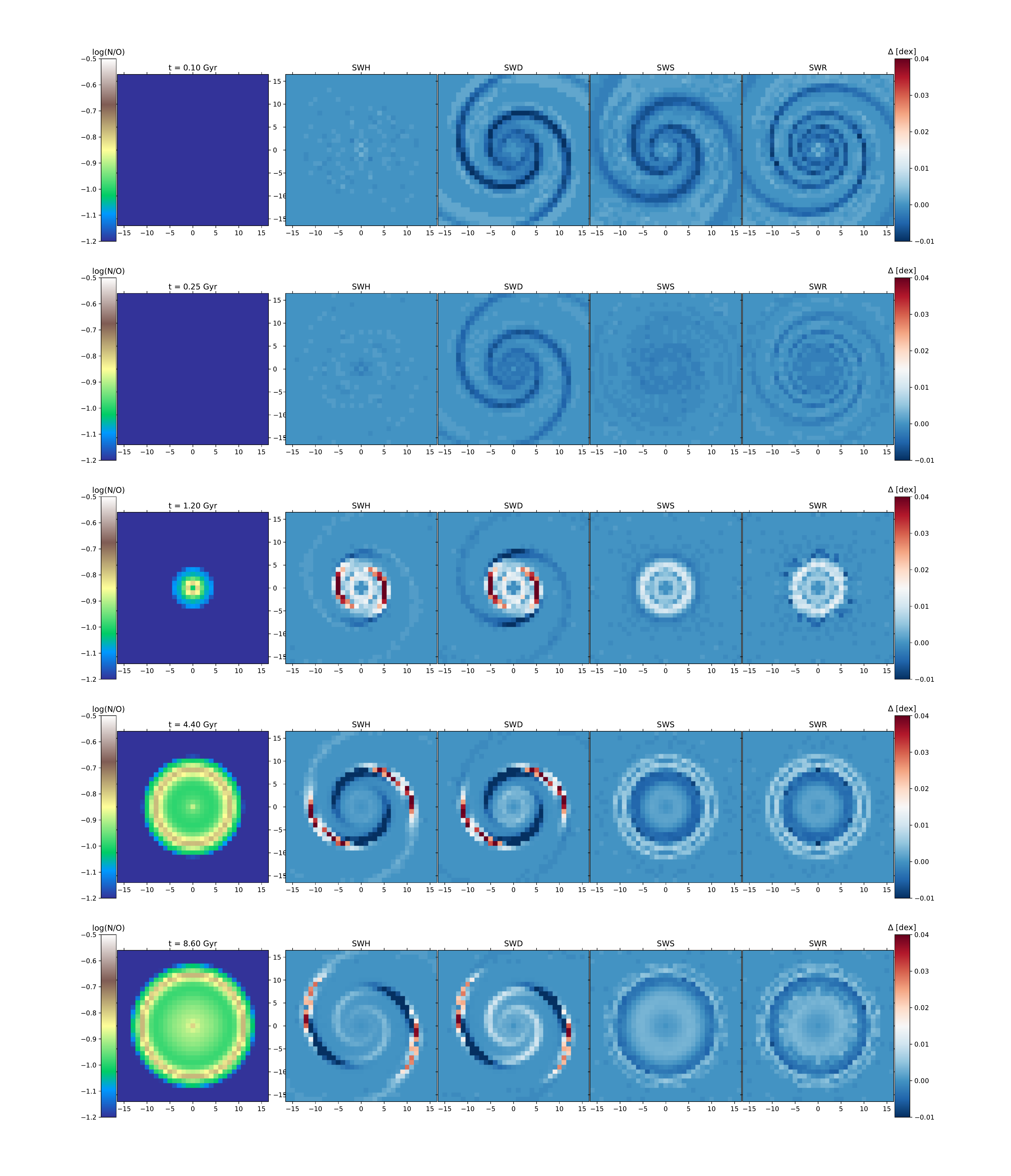}
\vspace{-1cm}
\caption{Time evolution of the nitrogen-to-oxygen ratio $\log{(N/O)}(x,y)$ at: top)  $t=0.1$\,Gyr; second row) $t=0.25$\,Gyr; third row) $t=1.20$\,Gyr; fourth row) $t=4.40$\,Gyr; and bottom) $t=8.60$\,Gyr. Left panels correspond to the AZ model, with scale at the left; while the right columns are the differences of SW models compared with AZ results: second column SWH-AZ; third column, SWD-AZ, and fourth column SWR-AZ.  In these three cases we represent differences with the same color code: from blue (negative differences) to red (the positive ones).}
\label{NO_t_2D}
\end{figure*}

Finally, we plot the nitrogen-to-oxygen abundance ratio $\log{(N/O)}$ in Figure~\ref{NO_t_2D}.
In this case, the residuals are small, but, interestingly, they become negative in some regions for all models, mainly in the outer part of the spiral arms, where other quantities reach their maxima.

One may conclude from the above results that differences between models AZ and SWH, where the overdensity is assumed to be present in the halo at the initial timestep, are not large. In particular, for the present time, they are not larger than 0.03\,dex.
Such a value lies below the typical uncertainties in the determination of chemical abundances (and abundance ratios) from observational data.
Only for the SFR, measured via H$\alpha$ intensity, would it be (potentially) possible to see such small differences. It is also clear that if we want to search for residuals larger than the observational uncertainties, it would be necessary to use the results for the first times after the creation of the spiral arm since its effects will erase rapidly after 1-2\,Gyr. 


\begin{figure*}
\includegraphics[width=0.49\textwidth,angle=0]{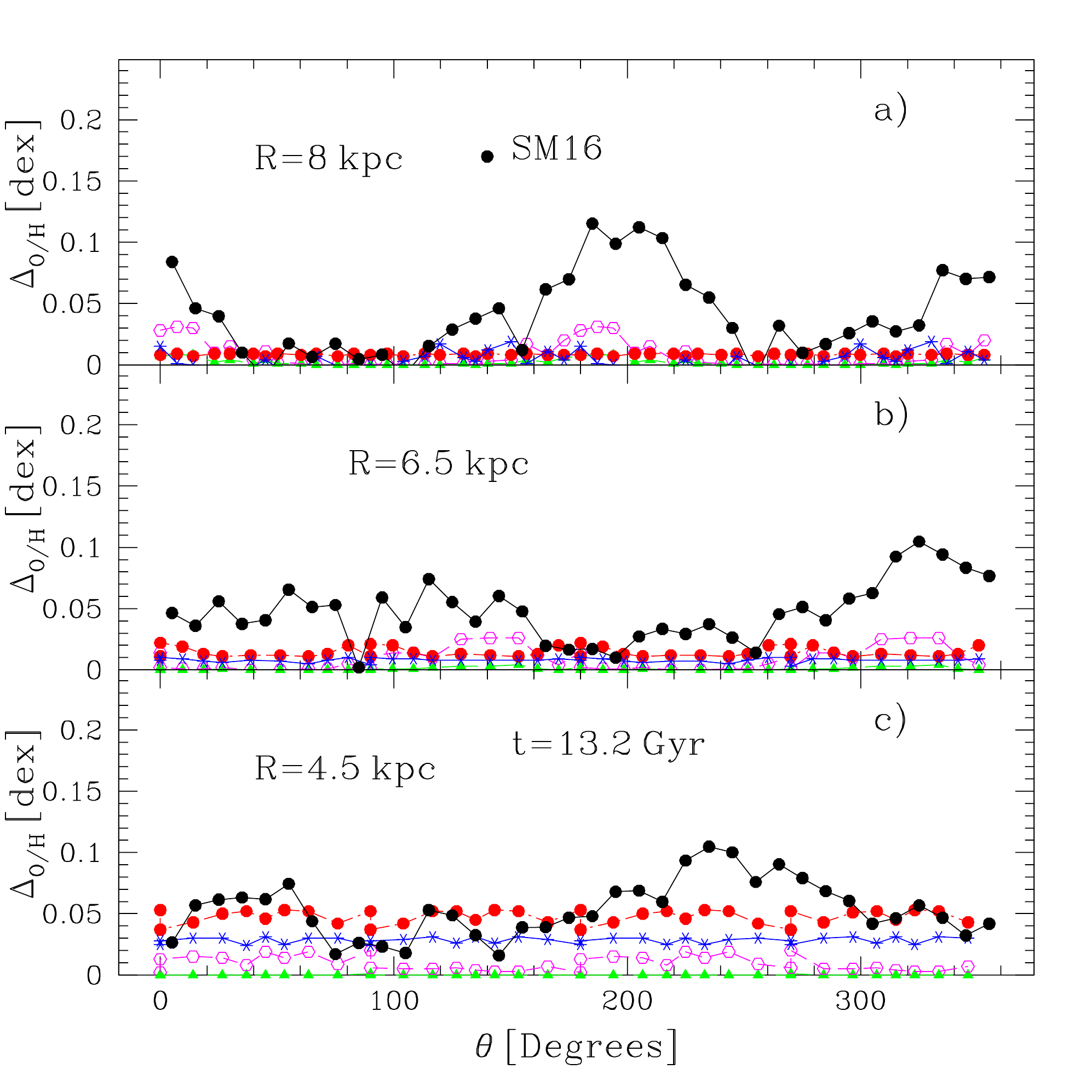}
\includegraphics[width=0.49\textwidth,angle=0]{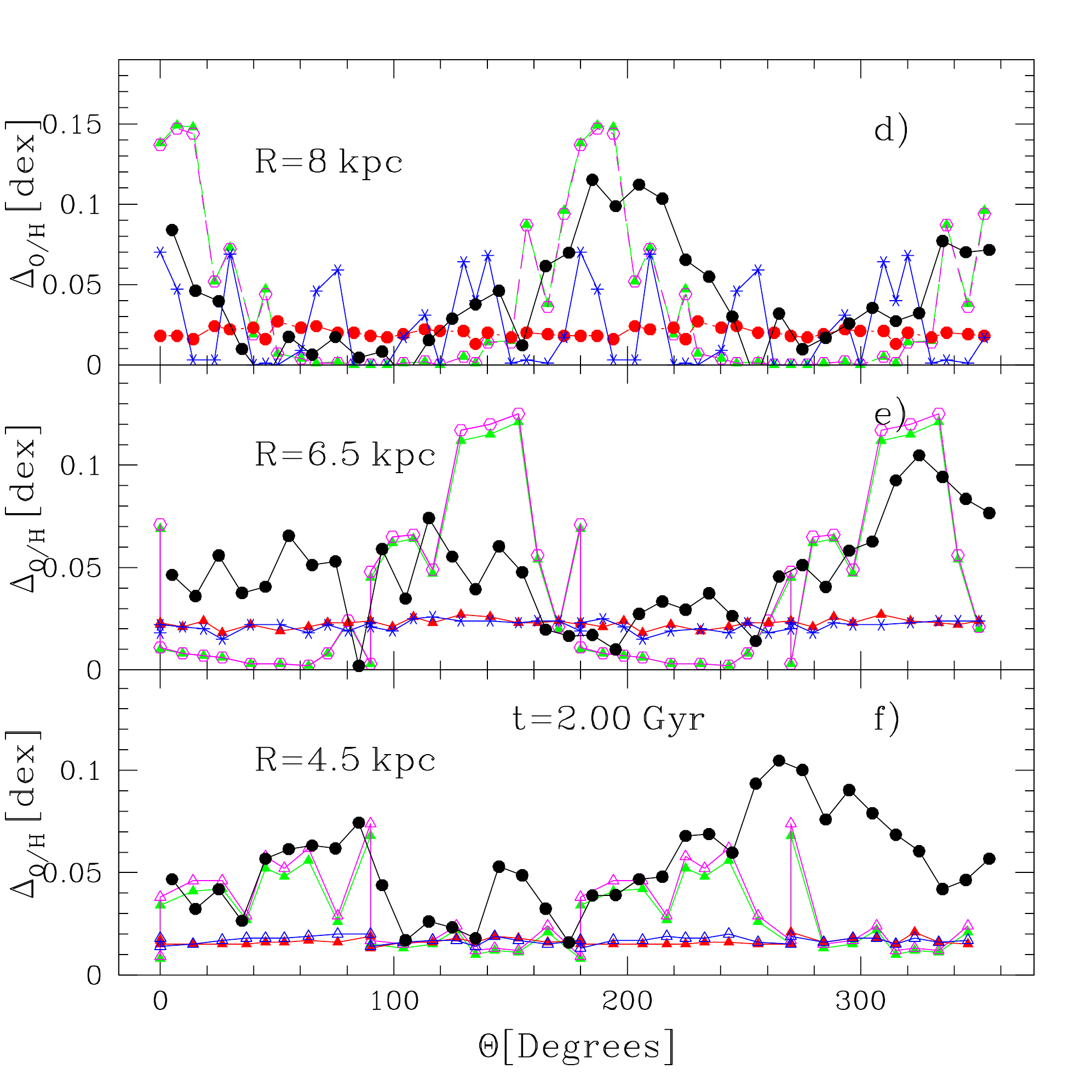}
\caption{Differences in oxygen abundances $12+log(O/H)$ between AZ and SWH,SWD, SWS and SWR models --in cyan, green, red and blue lines, respectively-- as a function of the angle $\theta$ for three different galactocentric radii as labelled, compared with the data from SM16 for oxygen abundances differences between arm and inter-arm regions --solid black dots and line. Left panels correspond to $t=13.2$\,Gyr,  while the right ones refer to $t=2.0$\,Gyr.}
\label{sim}
\end{figure*} 

This is shown with the comparison of our results with observations obtained with VLT/MUSE for the galaxy NGC~6754, as detailed in \citet{sanmen16b}.  We show oxygen abundance residuals as a function of the azimuthal angle, $\theta$, in Figure~\ref{sim}, at three different galactocentric distances. Thus, in a), we have regions located in $R\sim 8$\,kpc, in b), $R\sim 6.5$\,kpc and in c), $R\sim 4$\,kpc, for the present time. Taking into account that the galaxy is not the  MWG, we would not expect the model to fit exactly the observations, since the spiral density wave might have a different contrast, angular pattern, and pitch angle as those assumed here. For this reason, we have moved the angle of observations by a given constant quantity in each panel. Nevertheless, t is still impossible to reproduce these data with our model abundances for the present time. Therefore, taking into account our previous results, we have used the abundances at an alternative time, $t= 2.00$\,Gyr, in panels d) , e) and f),  at the same galactocentric distances as in the left panels. Our results are then in better agreement with the shape of the observations found in NGC~6754 \citep[see Figure 2 from][]{ sanmen16b} but only for the SWH and SWD models, including a similar decrease on both sides of each maximum. Probably for SWS and SWR it would be necessary to select a time even earlier.
The characteristic pattern clearly seen in the right panels --see green, magenta and black lines in panel f)--, with an increase and then a decrease around the typical angle of the spiral arm, is similar to the expected behavior seen by \citet{ho19} (see his Figure\,p23), obtained from \citet{ho17,ho18} observations (represented as a function of the polar angle). So, we may reproduce this type of variation, including residuals in the appropriate range, with the SWH and SWD models, if we use the early times results.

\section{Conclusions}
\label{con}

A multitude of studies over the past 50 years have explored the chemical evolution of spirals within a 1D framework. In this work, a new generation of 2D models has been developed, building upon our own 1D base.

We have run a chemical evolution model with azimuthal symmetry (AZ), which may be compared with the present time radial distributions of stellar density profile, SFR, O/H and N/O abundances, as well as four other models including a spiral density wave (SW) as an overdensity added to the AZ disc. These SW models have been compared with the AZ one, in order to estimate effects of the SW for a spiral galaxy similar to MWG.

The main findings of this study are as follows:
\begin{enumerate}

\item The over-density of the spiral wave modifies only slightly the stellar profile at the present time, the SW models showing differences with AZ as small as $\sim 0.05$\,dex. Models SWH and SWD without variation of the SW with time, show higher residuals in the outer regions of the disc ($R>15$\,kpc). 

\item For the SFR, models show differences for the present time both in the inner and the outer regions of the disc. The outer regions are more affected by the spiral arm in models SWH and SWD. If the rotation of the spiral density wave is taken into account, in models SWS and SWR, the inner regions of the disc have larger differences in SFR compared to the outer ones, reaching values as large as 0.2\,dex in a relatively narrow ring around 5\,kpc of galactocentric distance,showing a light dependence on azimuthal angle.

\item Changes in the distribution of oxygen abundances, 12+log(O/H),
  at the present time due to the spiral wave are smaller
  ($\sim$0.03~dex) than in the SFR, and they are below the typical
  empirical uncertainties associated with observational abundance
  determinations. They are slightly higher in the outer regions of the
  disc where reach values of 0.1\,dex, 
The effect of the SW on the N/O abundance ratio is even more difficult to detect, although a abrupt variation appears around 100 and 300$^{\degree}$ at $R= 14-15$\,kpc, with a decrease just before and an increase just after the spiral arm, with residuals of $\sim 0.02-0.03$, in the limit of observational uncertainties.

\item The differences between AZ and the other four models including the spiral wave, as well as the contrast between the arm and inter-arm regions, change significantly with time, being stronger shortly after including the SW.
Furthermore, the azimuthal differences in the last two models including rotation (SWS and SWD) dilute very quickly as a consequence of the high number of times ($\sim$50) that the spiral wave crosses every cell of the disc in our simulations. For evolutionary times greater than a few Gyr, azimuthal trends become hardly observable.

\item For earlier times, the contrast between arm and inter-arm regions may be similar to the observed one. This implies that if abundance differences arm--interarm there exist, the spiral arm had to be created very close in the time, that is, it last for 1-2\,Gyr before the observations, suggesting recurrent spiral wave along the evolution of a disc.

\end{enumerate}

In summary, the spiral arms cause an over-density that modifies only slightly the chemical abundances.
In models SWH and SWD, this over-density affects some regions of the galactic disc, mainly in the SFR, where differences are larger or of the same order of magnitude as the observational uncertainties.
It would be more challenging to see them in the elemental abundance maps, since they are mostly smaller than the classical uncertainties of  $\sim$0.10\,dex.
Models SWS and SWR show residuals in all quantities, compared with AZ, but the azimuthal pattern tends to dilute very rapidly with time.
This azimuthal pattern might be still visible in the SFR (or H$_{\alpha}$); it would be more challenging to observe in the abundances, but might be feasible in the case of a more recent SW where the arm vs inter-arm differences might be measurable.

If we compare the results with existing data, models with the SW included at late times may more easily reproduce the observations, which could imply that spiral arms are relatively recent features, $\sim 1-2$\,Gyr old or even younger.
With older ages, the effect on abundances would be diluted.
The possible existence of recurrent spiral density waves may be the reason for the observed differences.
We shall revisit this subject in our next paper, in which we shall consider variations of the parameters defining the SW and explore alternative formulations that provide a more realistic description of the formation and evolution of the spiral density wave.

\section{Acknowledgments}
The authors acknowledge the anonymous referee for their comments, which have dramatically improved the
manuscript.
This work has been partially supported by MINECO-FEDER-grants 
AYA2013-47742-C4-4-P, AYA2016-79724-C4-1-P and AYA2016-79724-C4-3-P. YA was supported by contract  RyC-2011-09461 of the \emph{Ram{\'o}n y Cajal} program. BKG acknowledges
the support of STFC through the University of Hull Consolidated Grant 
ST/R000840/1, and access to {\sc viper}, the University of Hull High 
Performance Computing Facility. This research was supported in part by the 
National Science Foundation under Grant No. PHY-1430152 (JINA Center for the 
Evolution of the Elements). OC acknowledges the support of FAPEMIG Grant APQ-00915-18.

\label{lastpage}
\end{document}